\documentclass[oldversion]{aa}  
\usepackage{graphicx,amsmath}
\usepackage{amssymb}
\usepackage{color}
\usepackage{natbib}		
\usepackage{url}
\usepackage{multirow}
\usepackage{commath}
\bibpunct{(}{)}{;}{a}{}{,} 

\def\approxsup{%
  \def\p{%
    \setbox0=\vbox{\hbox{$>$}}%
    \ht0=0.6ex \box0 }%
  \def\s{%
    \vbox{\hbox{$\sim$}}%
  }%
  \mathrel{\raisebox{0.7ex}{%
      \mbox{$\underset{\s}{\p}$}%
    }}%
}

\begin{document}

   \title{SOPHIE velocimetry of Kepler transit candidates XVI. Tomographic measurement of the low obliquity of KOI-12b, a warm Jupiter transiting a fast rotator.}

   \author{
   V.~Bourrier\inst{1,2}\and
   A.~Lecavelier des Etangs\inst{1}\and
   G.~H\'ebrard\inst{1,3} \and 
	 A.~Santerne \inst{4,5} \and 
	 M. Deleuil\inst{4} \and
	 J.M.~Almenara\inst{6} \and
	 S.C.C.~Barros\inst{4} \and
	 I.~Boisse\inst{4} \and
	 A.S.~Bonomo\inst{7} \and
	 G.~Bruno\inst{4} \and
	 B.~Courcol\inst{4} \and
	 R.F.~Diaz\inst{2} \and
	 G.~Montagnier\inst{1,3} \and
	 C.~Moutou\inst{4,8}
   }

\institute{
Institut d'astrophysique de Paris, UMR7095 CNRS, Universit\'e Pierre \& Marie Curie, 98bis boulevard Arago, 75014 Paris, France 
\and 
Observatoire de l'Universit\'e de Gen\`eve, 51 chemin des Maillettes, 1290 Sauverny, Switzerland
\and
Observatoire de Haute-Provence, CNRS/OAMP, 04870 Saint-Michel-l'Observatoire, France
\and
Aix Marseille Universit\'e, CNRS, LAM (Laboratoire d'Astrophysique de Marseille) UMR 7326, 13388 Marseille, France
\and
Instituto de Astrof\'isica e Ci\^{e}ncias do Espa\c co, Universidade do Porto, CAUP, Rua das Estrelas, PT4150-762 Porto, Portugal
\and
UJF-Grenoble 1 / CNRS-INSU, Institut de Plan\'etologie et d'Astrophysique de Grenoble (IPAG), UMR 5274, 38041 Grenoble, France
\and
INAF - Osservatorio Astrofisico di Torino, Via Osservatorio 20, 10025 Pino Torinese, Italy
\and
Canada-France-Hawaii Telescope Corporation, 65-1238 Mamalahoa Hwy, Kamuela, HI 96743, USA
}

\authorrunning{V.~Bourrier et al.}
\titlerunning{Tomographic measurement of KOI-12b}

\offprints{V.B. (\email{vincent.bourrier@unige.ch})}  

   \date{}
 
  \abstract
{
We present the detection and characterization of the transiting warm Jupiter KOI-12b, first identified with Kepler with an orbital period of 17.86 days. We combine the analysis of Kepler photometry with Doppler spectroscopy and line-profile tomography of time-series spectra obtained with the SOPHIE spectrograph to establish its planetary nature and derive its properties. To derive reliable estimates for the uncertainties on the tomographic model parameters, we devised an empirical method to calculate statistically independent error bars on the time-series spectra. KOI-12b has a radius of 1.43$\pm$0.13\,$R_\mathrm{Jup}$ and a 3$\sigma$ upper mass limit of 10$M_\mathrm{Jup}$. It orbits a fast-rotating star ($v$sin$i_{\star}$ = 60.0$\pm$0.9\,km\,s$^{-1}$) with mass and radius of 1.45$\pm$0.09\,$M_\mathrm{\sun}$ and 1.63$\pm$0.15\,$R_\mathrm{\sun}$, located at 426$\pm$40\,pc from the Earth. Doppler tomography allowed a higher precision on the obliquity to be reached by comparison with the analysis of the Rossiter-McLaughlin radial velocity anomaly, and we found that KOI-12b lies on a prograde, slightly misaligned orbit with a low sky-projected obliquity $\lambda$ = 12.6$\stackrel{+3.0}{_{-2.9}}^\circ$. The properties of this planetary system, with a 11.4 magnitude host-star, make of KOI-12b a precious target for future atmospheric characterization.\\

%
}


   \maketitle

\section{Introduction}
\label{intro} 

The transit of an exoplanet across its rotating host star distorts the apparent stellar line shape by removing the part of the profile emitted by the occulted portion of the star. This induces anomalous variations in the measured stellar radial velocity during the transit, known as the Rossiter-McLaughlin (RM) anomaly (\citealt{holt1893}; \citealt{rossiter1924}; \citealt{mclaughlin1924}). The shape of the anomaly as a function of time depends on the value of the sky-projected spin-orbit obliquity $\lambda$, which is the angle in the plane of the sky between the projection of the stellar spin axis and the projection of the orbital angular momentum vector. Most obliquity measurements have been obtained using Doppler spectroscopy, but complementary techniques also make use of spot-crossing events during planetary transits  (\citealt{Nutzman2012}; \citealt{SanchisOjeda2012}), gravity darkening (\citealt{szabo2011}; \citealt{barnes2013}), asteroseismology (\citealt{Chaplin2013}) and Doppler tomography (\citealt{cameron2010a}). \\
Obliquity is a good tracer of planetary system histories, as different formation scenarios result in different spin-orbit angles. About thirty misaligned systems ( $\abs{\lambda}>$30$^{\circ}$ and inconsistent with $\lambda=$0$^{\circ}$) have been identified today over more than eighty measured systems\footnote{\mbox{the Holt-Rossiter-McLaughlin Encyclopaedia:} \mbox{\url{http://www.physics.mcmaster.ca/~rheller/}}} (\citealt{albrecht2012}; \citealt{Crida2014}), including some with retrograde or nearly polar orbits (e.g. \citealt{winn2009c}; \citealt{narita2010}; \citealt{triaud2010}; \citealt{hebrard2011}). Most of these measurements have been done for close-in, isolated giant planets. While it is commonly accepted that hot-Jupiters form beyond the snow line and later migrate toward the star, many unknowns remain about how the migration occurs. Their wide distribution of obliquities favours misaligning scenarios where massive giant planets have been brought in by planet-planet (or planet-star) scattering,
and Kozai migration with tidal friction (see e.g. \citealt{fabrycky2007}; \citealt{guillochon2011}, \citealt{Naoz2011}). Some models show that the initial misalignment of a planet could also be maintained through its interactions with the disk (\citealt{Teyssandier2013}). Instead, more standard scenarios implying disk migration are expected to conserve the initial alignment between the angular momentums of the disk and of the planetary orbits (see, e.g., \citealt{Lin1996}). Alternatively, it has been proposed that the orbit still reflects the orientation of the disk, with the stellar spin instead having moved away, either early-on through magnetosphere-disk interactions (\citealt{lai2011}) or later through elliptical tidal instability (\citealt{cebron2011}). To understand whether such scenarios and the resulting wide distribution of obliquities are specific to massive close-in exoplanets, it is necessary to sample all types of planetary systems. While obliquity measurements have now extended to smaller planets (e.g. \citealt{winn2010b}; \citealt{Hirano2012}; \citealt{albrecht2013}; \citealt{bourrier2014b}; \citealt{lopez2014}), they remain little known for long-period exoplanets (five cases\footnote{Kepler-30 d, Kepler-30 c, Kepler-30 b, \citet{SanchisOjeda2012}; HD\,80606 b,	\citet{hebrard2010}; HD\,17156 b, \citet{narita2008}} with $P\approxsup$11\,days) or those orbiting fast-rotating stars (five cases\footnote{HAT-P-32 b and HAT-P-2 b, \citet{albrecht2012}; CoRoT-11 b, \citet{ Gandolfi2012}; WASP-33 b, \citet{cameron2010}; Kelt-7b, \citet{Bieryla2015}} with $v$sin$i_{\star}\approxsup$20\,km\,s$^{-1}$). \\
The Kepler candidate KOI-12.01 (Table~\ref{tab_info}) offers the opportunity to probe spin-orbit angles in both domains, as it orbits a moderately bright fast-rotator (Kepler magnitude 11.353; $v$sin$i_{\star}\sim$66\,km\,s$^{-1}$, \citealt{Santerne2012}) with a period of $\sim$18 days. Detected with the Kepler satellite (\citealt{Batalha2013}), KOI-12.01 was first studied by \citet{Demory_seager2011} as a potentially inflated planet. Unfortunately, the authors were not able to conclude on the planetary nature of the transit signal based on Kepler photometry alone. Binary systems can mimic a planetary transit signal and are the sources for a non negligible part of the Kepler candidates (e.g. \citealt{Santerne2012}; \citealt{Fressin2013}). This is particularly true for candidates showing transit depths around 1\% and thus possibly corresponding to giant planets, which is the case for KOI-12.01. \citet{Santerne2012} further investigated the nature of KOI-12.01 using radial velocity follow-up, but it proved unsufficient to solve the candidate. In recent years, Doppler tomography has been used to study the alignement and properties of a growing number of planetary systems (HD\,189733, \citealt{cameron2010a}; WASP-3, \citealt{miller2010}; CoRoT-11b, \citealt{Gandolfi2012}; WASP-32, -38, -40, \citealt{Brown2012}; Kelt-7b, \citealt{Bieryla2015}) but also to assess the planetary nature of the transiting gas giant WASP-33b (\citealt{cameron2010}). This technique, similar to the methodology initially developped to model the RM effect in binary stars (\citealt{Albrecht2007}, \citealt{Albrecht2009}) relies on the decomposition of the stellar line profile into its different components, namely the stellar and instrumental profile, the limb-darkened rotation profile, and the travelling signature caused by the transiting planet. The large rotational broadening of the stellar lines for stars like WASP-33 and KOI-12 introduces errors when fitting the radial velocities of the RM anomaly with analytical formulae (e.g. \citealt{Triaud2009}, \citealt{Hirano2010}). In contrast, Doppler tomography is best-suited to the analysis of such systems, as it models directly the missing starlight signature caused by the planet and reponsible for the radial velocity anomaly.\\
Here we combine Kepler photometry and SOPHIE time-series spectroscopy to assess the nature of KOI-12.01 and determine the planetary system properties. Observations and data reduction are described in Sect.~\ref{obs}. The analysis of Kepler photometry is presented in Sect.~\ref{photom}. Radial velocities are analyzed in Sect.~\ref{RV_section}, and spectroscopic data are also used in Sect~\ref{doppler} to perform line-profile tomography. The planetary nature of KOI-12.01 is discussed in Sect.~\ref{planet}, and we summarize our results in Sect.~\ref{conclu}.

\begin{table}[tbh]
  \caption{IDs, coordinates and magnitudes of KOI-12.}
  \label{tab_info}
\begin{tabular}{lc}
\hline
Parameter & Value \\
\hline
Kepler object of interest     &    KOI-12  \\
Kepler exoplanet catalog      &    Kepler-? \\
Kepler Input catalog			 & KIC 5812701   \\
2MASS ID						&  19494889+4100395  \\
WISE ID						&  194948.89+410039.6 \\
\hline
RA (2000.0) & 19$^{h}$49$^{mn}$48$^{s}$.90 \\
Dec. (2000.0) & +41$^{\circ}$0$^{'}$39$^{''}$.56 \\
\hline
Kepler magnitude & 11.353 \\ 
SDSS-G & 11.571 \\ 
SDSS-R & 11.280 \\ 
SDSS-I & 11.245 \\ 
2MASS-J & 10.461$\pm$0.020 \\ 
2MASS-H & 10.287$\pm$0.022 \\ 
2MASS-K$_{s}$ & 10.234$\pm$0.018 \\ 
WISE-W1 & 10.189$\pm$0.023 \\ 
WISE-W2 & 10.198$\pm$0.020 \\ 
WISE-W3  &  10.015$\pm$0.042 \\
\hline
\end{tabular}
\end{table}


\section{Observations and data reduction}
\label{obs} 

\subsection{Photometric data with Kepler}
\label{phot_obs} 

\begin{figure}[tbh]	
\centering
\includegraphics[trim=0cm 0cm 0cm 0cm, clip=true,width=0.9\columnwidth]{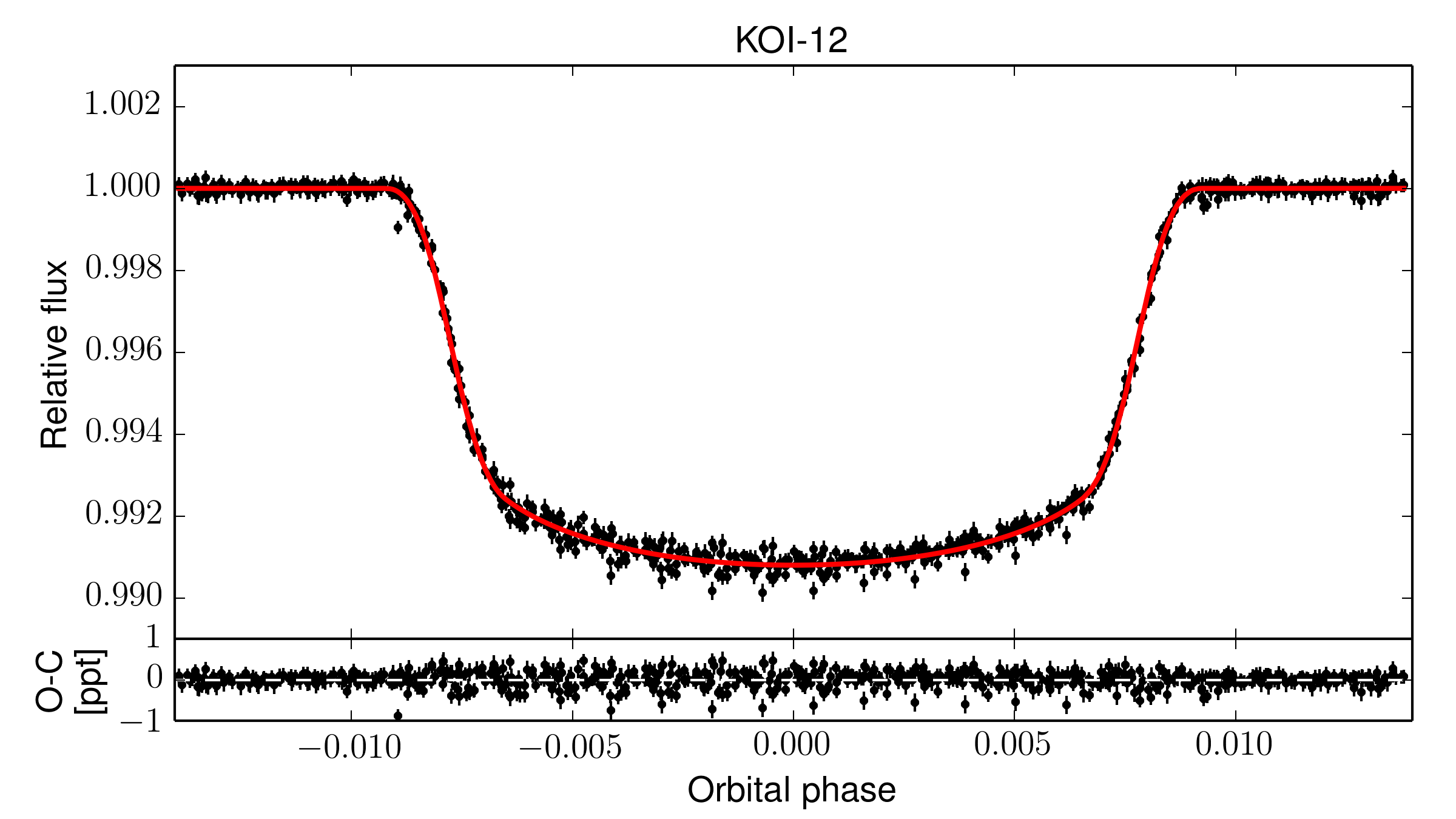}	
\caption[]{Long-cadence transit light curve of KOI-12.01, normalized and phase-folded over the Kepler observations. The best fit to the data is displayed as a red line, with residuals in the lower panel. One of the light curves was found to be abnormally deep, and there is also a small increase in the residual scatter during the transits. These variations should not affect the derived parameters, but may slightly increase their uncertainties.}
\label{fig:lc} 
\end{figure}

Seventy-four transits of the planetary candidate KOI-12.01 were observed with Kepler (\citealt{Batalha2011,Batalha2013}), with a period of 17.9 days and transit depth of about 1\,\%. No transits with different periods were detected in any of the light curves, and there are thus no signs that KOI-12 is a multiple transiting system. The Kepler photometry was acquired in long-cadence (LC) and short-cadence (SC) data, with one point per 29.42~minutes and 58~seconds, respectively. We used in the present analysis a combination of data in both cadences, with 28 LC transits and 46 SC transits (Table~\ref{table_log_kep}). We used the light-curve of quarters Q0 to Q17 reduced by the Photometric Aperture Kepler pipeline, that accounts for barycentric, cosmic ray, background, and so-called argabrightening corrections (\citealt{Jenkins2010}), publicly available from the MAST archive (http://archive.stsci.edu/kepler).\\
In addition to the transits, the Kepler light curve shows quasi-periodic photometric variability with typical amplitudes of 1000\,ppm on timescales similar to the transit duration ($\sim$7.5\,h). They are probably due to inhomogeneities of the stellar surface (spots, plages, etc) modulated with the rotation of the star. Before modeling the transits, we normalized fragments of the light curves by fitting an out-of-transit 2$^{nd}$ order polynomial. Because of the relatively short timescale of the stellar variability, we limited this fit to $\sim$1.9\,hours of out-of-transit data (corresponding to more than 200 photometric measurements in the SC data and about 8 in the LC data). Given the high quality of the Kepler LC photometry, we found that this was sufficient to detrend all but one of the transits. While we removed most of the stellar variability with a timescale greater than the transit duration, stellar variability with a shorter timescale is not corrected and may explain the small increase in the scatter of the photometric data observed during the transit in Fig.~\ref{fig:lc}. Planet-spot occultations might also explain this increased scatter, which does not affect significantly the derived parameters but may slightly increase their uncertainties. Since the Kepler spacecraft rotates four times a year, the crowding values are different between seasons. We thus produced four crowding-uncorrected de-trended light curves for each season, using the crowding estimates provided by the MAST database (given the relative brightness of the target, we estimated an uncertainty of 0.2\%). This allowed us to account for differential crowding values, noises, and out-of-transit fluxes in the transit modeling. Fig.~\ref{fig:lc} shows the corresponding phase-folded transit light curves, once normalized and corrected for crowding.

\begin{table}[th]
  \caption{Mid-transit epoch of the LC and SC Kepler light-curves of KOI-12 used in our photometric analysis.}
  \label{table_log_kep}
\begin{tabular}{cc}
\hline
\hline
\multicolumn{2}{c}{BJD-2400000}   \\
    Short cadence &  Long cadence 	   \\
\hline
55033.16169				& 54979.59599 \\
55051.01692				& 54997.45122 \\
55104.58262				& 55068.87216  \\
55122.43786				& 55086.72739  \\
55140.29309				& 55158.14832  \\
55193.85879				& 55176.00356  \\
55211.71402				& 55247.42449  \\
55229.56926 				& 55265.27972  \\	
55283.13496 				& 55354.55589  \\
55300.99019 & 55443.83206  \\
55318.84542 & 55461.68729  \\
55336.70066 & 55533.10822  \\
55390.26636 & 55550.96346  \\
55408.12159 & 55622.38439  \\
55425.97682 & 55711.66056  \\	
55479.54252 & 55729.51579  \\
55497.39776 & 55818.79196  \\
55515.25299 & 55908.06812  \\				
55568.81869 & 55925.92335  \\
55586.67392 & 55997.34429  \\
55604.52916 & 56086.62045  \\			
55658.09486 & 56104.47569  \\
55675.95009 & 56175.89662  \\
55693.80532 & 56193.75185  \\	
55747.37102 & 56283.02802  \\
55765.22626 & 56300.88325  \\
55783.08149 & 56372.30419  \\
55800.93672 & 56390.15942  \\	
55836.64719 & \\
55854.50242 & \\
55872.35766 & \\
55890.21289 & \\	
55943.77859 & \\
55961.63382 & \\
55979.48905 & \\		
56033.05475 & \\
56050.90999 & \\
56068.76522 & \\		
56140.18615 & \\
56158.04139 & \\			
56211.60709 & \\
56229.46232 & \\
56265.17279 & \\				
56336.59372 & \\
56354.44895 & \\				
56408.01465 &   \\
\hline
\multicolumn{2}{l}{Notes: Kepler transit epochs}\\
\multicolumn{2}{l}{are given in Barycentric Dynamical Time.}\\
\end{tabular}
\end{table}

\subsection{Spectroscopic data with SOPHIE}
\label{spec_obs} 

Spectroscopic measurements of KOI-12 were made with the spectrograph SOPHIE at the 1.93-m telescope of Haute-Provence Observatory (France). Two transits of KOI-12.01 were observed in June 2012 (27 measurements, 20 during the transit) and June 2013 (18 measurements, 12 during the transit), in order to detect the RM anomaly and perform line-profile tomography. Additional measurements were taken far from the transit in 2011 (2 in February and June 2011; \citealt{Santerne2012}) and in 2014 (2 in May and June 2014) to put constraints on the mass of KOI-12.01. The 2014 observations are of better quality than the 2011 ones, which may be affected by a systematic shift reminiscent of previous SOPHIE observations in High-Efficiency mode (\citealt{hebrard2013}). Contrary to the 2014 dataset, no radial velocity standard stars were monitored during the 2011 observations, and we could not find any adequate correction for this possible shift. We thus used only observations taken in 2012, 2013 and 2014.\\
The transit of KOI-12.01 lasts 7.5 hours, which makes difficult the observation in good conditions of the whole event during a night with a ground-based telescope. Run 2012 was executed over 3 nights: the second night offers a good coverage of the whole transit, while reference measurements were taken outside the transit during the previous and the following nights. Run 2013 was executed over 2 nights: only the first part of the transit could be observed during the first night with partial coverage, although reference observations were also secured immediately before the transit, with an additional measurement the following night. Run 2014 is made of two measurements taken a few days apart, close to the quadrature phases, to constrain the Keplerian semi-amplitude. Run 2013 was obtained in poor weather conditions and provide data of lower accuracy than Runs 2012 and 2014. The log of the transit observations is reported in Table~\ref{table_log}.    \\
SOPHIE is a cross-dispersed, environmentally stabilized echelle spectrograph dedicated to high-precision radial velocity measurements (\citealt{Bouchy2009b}). The light is dispersed on 39 spectral orders from 387 to 694\,nm (labelled 0 to 38 from blue to red wavelengths). SOPHIE data were acquired in High-Efficiency mode (resolution power $\lambda/\Delta\lambda$ = 40000) with exposures ranging from 600 to 1800 seconds depending on weather conditions, to ensure a signal-to-noise ratio per pixel at 550\,nm as constant as possible (S/N$_\mathrm{550}\sim$35). Spectra were extracted using the standard DRS SOPHIE pipeline, and passed through weighted cross-correlation with a numerical mask (\citealt{baranne1996}, \citealt{pepe2002}). The pipeline does not normalize the spectra prior to computing the cross-correlation in order to keep a natural weighting by the flux. The resulting cross-correlation functions (CCF) are fitted with Gaussians to retrieve the radial velocities (Sect.~\ref{RV_section}), or fitted with a line-profile model to perform tomography (Sect.~\ref{doppler}). We found that the quality of these fits was improved when removing some low-S/N spectral orders from the cross-correlation. As expected from the effects of atmospheric dispersion, blue spectral orders yielded the lowest S/Ns and CCFs were calculated using spectral orders 5 -- 38 for Runs 2012 and 6 -- 38 for Run 2013. We discarded the fourth measurement in dataset 2013 because of its particularly low quality (S/N$_\mathrm{550}$=15). We tried different kinds of numerical masks when performing cross-correlation, and although KOI-12 is more akin to a F-type star, the lowest dispersion on the order-per-order RVs was obtained using a standard G2-type mask. Finally, the solar light reflected by the Moon can distort the shape of the CCF and shift the measured radial velocity. Moonlight contamination was detected in 3 exposures in Run 2012 and 6 exposures in Run 2013, and was corrected for by using the second fiber aperture targeted on the sky (\citealt{pollacco2008}; \citealt{hebrard2008}). We selected 45 SOPHIE spectra with a S/N at 550\,nm greater than 20 to build a combined 1D spectrum to be used for the spectroscopic analysis of the host star (Sect~\ref{spectralana}). Individual exposures were set in the rest frame then co-added order per order using weights proportional to the S/N. The resulting co-added spectrum has a high S/N in the continuum of 150 per pixel at 550 nm (corresponding to a S/N of 548 per resolution element). \\
The SOPHIE pipeline may yield uncorrect estimates on radial velocities error bars when the fitted CCFs are affected by large rotational broadening. Since KOI-12 is a fast-rotator, we used instead an empirical formula well suited to SOPHIE radial velocity measurements, with error bars defined as 3.4\,$\sqrt{FWHM_{CCF}}$/(C\,x\,S/N$_\mathrm{550}$), where $FWHM_{CCF}$ is the full-width at half-maximum of a given CCF and $C$ its contrast defined as the relative difference between the CCF minimum and the continuum. The radial velocity measurements and their uncertainties are reported in Table~\ref{table_log_RV}. Error bars on the CCFs points, needed in the tomography analysis, were estimated from the dispersion of the residuals between the CCFs and their best-fit line-profile model (Section~\ref{noise}).

\begin{table*}[tbh]
  \caption{Log of KOI-12.01 transit observations with SOPHIE.}
  \label{table_log}
\begin{tabular}{lcccccc}
\hline
Run & Transit mid-time (UT) & Exposures$^\dagger$  & S/N\_450$^\ast$ & S/N\_550$^\ast$ & S/N\_650$^\ast$ \\
\hline
2012 & 2012-06-25 at 23h28 & 20 &  19 & 34 & 35 \\
2013 & 2013-06-18 at 01h58 & 12 &  18 & 34 & 35 \\
\hline
\multicolumn{6}{l}{$\dagger$: number of individual exposures taken during the transit.} \\
\multicolumn{6}{l}{$\ast$: medians of the signal-to-noise ratio per pixel during the transit at 450\,\AA, 550\,\AA, and 650\,\AA\.} \\ 
\end{tabular}
\end{table*}

\begin{table}[th]
  \caption{Radial velocities of KOI 12.}
  \label{table_log_RV}
\begin{tabular}{lccc}
\hline
\hline
			& BJD (UTC) 		 & RV 			& $\pm$\,1\,$\sigma$		\\
			& -2400000 & (km\,s$^{-1})$ & (km\,s$^{-1})$        \\
\hline
Run 2012   & 56103.5518& -21.87& 0.48\\
      & 56104.3650& -21.26& 0.48\\
      & 56104.3832& -20.95& 0.40\\
      & 56104.4030& -20.88& 0.42\\
      & 56104.4201& -21.86& 0.39\\
      & 56104.4381& -21.96& 0.34\\
      & 56104.4580& -22.29& 0.40\\
      & 56104.4766& -23.00& 0.39\\
      & 56104.4884& -22.57& 0.38\\
      & 56104.4999& -22.76& 0.39\\
      & 56104.5110& -21.90& 0.37\\
      & 56104.5208& -22.11& 0.38\\
      & 56104.5299& -22.85& 0.36\\
      & 56104.5387& -22.98& 0.38\\
      & 56104.5480& -22.65& 0.37\\
      & 56104.5587& -22.80& 0.36\\
      & 56104.5675& -22.39& 0.37\\
      & 56104.5760& -22.76& 0.38\\
      & 56104.5861& -22.22& 0.39\\
      & 56104.5982& -22.38& 0.37\\
      & 56104.6155& -22.60& 0.40\\
      & 56105.4795& -22.26& 0.36\\
      & 56105.4886& -22.07& 0.38\\
      & 56105.4973& -21.99& 0.38\\
      & 56105.5351& -22.31& 0.38\\
      & 56105.5453& -21.99& 0.38\\
      & 56105.5568& -22.60& 0.36\\
\hline
Run 2013         & 56461.3690& -24.76& 0.41\\
      & 56461.3842& -24.57& 0.37\\
      & 56461.3992& -24.35& 0.38\\
      & 56461.4126& -24.72& 0.36\\
      & 56461.4259& -25.88& 0.73\\
      & 56461.4409& -24.35& 0.39\\
      & 56461.4559& -23.00& 0.54\\
      & 56461.4710& -23.16& 0.51\\
      & 56461.5313& -24.25& 0.36\\
      & 56461.5401& -24.12& 0.37\\
      & 56461.5488& -24.29& 0.39\\
      & 56461.5575& -24.02& 0.38\\
      & 56461.5662& -24.55& 0.39\\
      & 56461.5747& -24.24& 0.39\\
      & 56461.5831& -23.93& 0.39\\
      & 56461.5915& -24.07& 0.39\\
      & 56461.6010& -24.36& 0.38\\
      & 56462.5911& -24.24& 0.69\\
\hline
Run 2014      & 56807.5694& -24.46& 0.23\\
      & 56814.4686& -24.39& 0.19\\
\hline
\multicolumn{4}{l}{Note: Variations in the asymmetric continuum of the}\\ 
\multicolumn{4}{l}{CCFs (Fig.~\ref{fig:CCFs}), instrumental effects, or the presence}\\
\multicolumn{4}{l}{of a massive distant companion to KOI-12 may cause}\\
\multicolumn{4}{l}{the differences in systemic velocities between the}  \\
\multicolumn{4}{l}{three runs}  \\
\end{tabular}
\end{table}

\section{Spectral analysis of the host star}
\label{spectralana}

\citet{Boisse2010} calibrated the relation between $v$sin$i_{\star}$ and the width $\sigma$ of the CCF in the high-efficiency (HE) mode of the SOPHIE spectrograph. The value for $\sigma$ depends on the B-V magnitude (0.22$\pm$0.12 for KOI-12; \citealt{hog2000}) and the full-width at half-maximum of the CCFs. The average value of FWHM$_{CCF}$ for the two runs is 87.5$\pm$1.9\,km\,s$^{-1}$ and yields $v$sin$i_{\star}$=60.0$\pm$1.5\,km\,s$^{-1}$. This calibration is however valid mostly for solar-type stars with a low or a moderate $v$sin$i_{\star}$. We thus checked the $v$sin$i_{\star}$ value by using the Fourier transform method (see \citealt{SimonDiaz2007}, and reference therein). From the first zero in the Fourier transform of various isolated spectral lines in the the co-added SOPHIE spectrum, we found $v$sin$i_{\star}$ = 66.0$\pm$2.5\,km\,s$^{-1}$ and $v_{macro}$=18.1$\pm$5.5\,km\,s$^{-1}$, consistent with the value $v$sin$i_{\star}$ = 70.0$\pm$5.0\,km\,s$^{-1}$ obtained by \citet{Lillobox2015}. The final value for the projected rotational velocity will be given in Sect.~\ref{results}, where it is determined more accurately using Doppler tomography.\\
 We derived the star's atmospheric parameters from the co-added SOPHIE spectrum of KOI-12 using the spectral analysis package SME 2.1 (\citealt{Valenti_Piskunov1996}; \citealt{valenti2005}). The fast rotation of the star prevents us to use alternative methods based on equivalent width measurements such as VWA. SME iteratively determines the fundamental stellar parameters by performing non-linear least-squares fit of synthetic spectra to the observed spectrum. The temperature was estimated on the extended wings of the Hydrogen Balmer lines. The other atmospheric parameters, $\log g_{\star}$, the metallicity, but also the various velocities, $v$sin$i_{\star}$, $v_{micro}$, and $v_{macro}$, were estimated on a large number of metal lines located in different spectral windows. We found $T_{\rm eff}$=6800$\pm$120\,K, $\log g_{\star}$=4.34$\pm$0.15, $[Fe/H]$=0.09$\pm$0.15, $v$sin$i_{\star}$=66.3$\pm$1.2\,km\,s$^{-1}$,  $v_{micro}$ =2.3$\pm$0.3\,km\,s$^{-1}$ and $v_{macro}$=17.3$\pm$1.4\,km\,s$^{-1}$. We note that the values we obtained for the projected velocity due to rotation, $v$sin$i_{\star}$, and the velocity due to convection, $v_{macro}$, are in agreement with the values we derived from the Fourier Transform of isolated spectral lines. From the values of the effective temperature, $\log g_{\star}$, and metallicity used as input values and the Starevol evolutionary tracks (\citealt{Lagarde2012}, Palacios, private communication) we derived the host-star's fundamental parameters, its mass $M_{\star}$ = 1.51$\stackrel{+0.14}{_{-0.09}}$\,M$_{\odot}$, radius $R_{\star}$=1.41$\pm$0.06$R_{\odot}$, and age 1.4$\pm$0.3\,Gyr.

\section{Photometry analysis}
\label{photom}

The normalized Kepler light curves were fitted by a transit model using the \texttt{PASTIS} code (\citealt{diaz2014}). The analysis also includes the fit of the SED and stellar evolution tracks to determine coherent stellar parameters. We used the Dartmouth (\citealt{dotter2008}) stellar evolution tracks as input for the stellar parameters.\\
For each Kepler light curve we included the out-of-transit flux and the contamination factors as free parameters. We also account for additional sources of Gaussian noise in the light curves and SED by fitting a jitter value to each dataset. This is especially appropriate for the Kepler data since the star is located on different CCDs each season. To account for the widening of the ingress and egress durations (\citealt{Kipping2010}), we modeled the LC light-curves using an oversampling factor of 10 suitable to the Kepler data (\citealt{Bonomo2014}). We fitted the free parameters using a Metropolis-Hasting Markov chain Monte Carlo (MCMC) algorithm (e.g. \citealt{Tegmark2004}) with an adaptive step size (\citealt{Ford2006}). To better sample the posterior distribution in the case of non-linear correlations between parameters, we applied an adaptive principal component analysis to the chains and jumped the parameters in an uncorrelated space (\citealt{diaz2014}). For most of the parameters of the MCMC we used non-informative priors (uniform or Jeffreys distributions). Exceptions are the stellar parameters $T_{\rm eff}$ and $[Fe/H]$ derived from the above spectral analysis, $\rho_{\star}$ obtained from $\log g_{\star}$ and the evolution tracks, and the orbital periods and phases of the planets for which we used as priors the Kepler values with error bars increased by a factor of 100 to be conservative. We used Gaussian priors for the contamination factors, centered on the MAST value (Sect.~\ref{phot_obs}) and with a width of 0.2\%. The priors on the stellar density were used to perform fits with eccentric orbits.\\
The system was analyzed with 10 chains leading to a total of 10$^{6}$ steps. Each chain was started at random points drawn from the joint prior. All chains converged to the same solution. We thinned the converged sub-chains using the correlation length. We finally merged the thinned chains, which left us with a total of more than 1000 independent samples of the posterior distribution. The resulting theoretical transit light curve is displayed in Fig.~\ref{fig:lc}. Because of an imperfect detrending or variations of the activity level of the star, one of the seventy-four transits was found to be abnormally deeper. Nonetheless, including it in the fit had no significant effect on the results. The best-fit parameters and their 68.3\,\% confidence intervals  are listed in Table~\ref{table:tab_paramsfit}. We did not detect a significant eccentricity, with a 3$\sigma$ upper limit of 0.72. Other parameter values were found to be in agreement with those derived for a circular fit, performed on the transit parameters without adjusting the stellar parameters (Table~\ref{table:tab_paramsfit}). Hereafter we choose parameters from the circular fits as final values. The results obtained in this section are also taken as final values for the stellar parameters. 

\begin{table*}[tbh]
\caption{Transit light curve and radial velocity analysis.}\centering
\begin{tabular}{lccl}
\hline
\noalign{\smallskip}  
\textbf{Parameter}    	& \textbf{Symbol} 			&  	    \textbf{Value}  		         	& \textbf{Unit} \\   
\noalign{\smallskip}
\hline
\hline
\noalign{\smallskip} 
Eccentric orbit     & & & \\
\hline
\noalign{\smallskip}
\textit{Orbital parameters} & & & \\
\hspace{0.4cm}Orbital period 		         & $P$ 					&		17.8552333$\pm$0.9$\times$10$^{-6}$			 & day \\
\hspace{0.4cm}Transit epoch 				 & $T_{0}$ 				&      	2454979.59599$\pm$2.7$\times$10$^{-4}$		 & BJD  \\
\hspace{0.4cm}Eccentricity                 & $e$                   &       $<$0.72\,(3$\sigma$)    								 & \\
\hspace{0.4cm}Argument of periastron       & $\omega$              &       228$\pm$120  								 & deg \\  
\hspace{0.4cm}Scaled semi-major axis 		& $a_\mathrm{p}/R_{\star}$     &	  20.0$\pm$1.5  & \\
\hspace{0.4cm}Semi-major axis & $a_\mathrm{p}$  & 0.151$\pm$3$\times$10$^{-3}$  & au \\
\hspace{0.4cm}Orbital inclination 				& $i_\mathrm{p}$ 				& 	88.95$\stackrel{+0.15}{_{-0.25}}$  		     & deg\\
\hspace{0.4cm}Impact parameter                  & $b$ 				&   0.3629$\pm$6.4$\times$10$^{-3}$                          &  \\
\hspace{0.4cm}Planet-to-star radii ratio 	& $R_\mathrm{p}/R_{\star}$ 	&		0.09049$\pm$8$\times$10$^{-5}$				   & \\
\hspace{0.4cm}Planet radius 	& $R_\mathrm{p}$ 	&	1.43$\pm$0.13	 & R$_{Jup}$\\
\hspace{0.4cm}Transit duration  & $t_\mathrm{d}$ &	6.788$\stackrel{+1.8\times10^{-2}}{_{-0.7\times10^{-2}}}$                          & hours \\
\hspace{0.4cm}Stellar reflex velocity   & $K$ 	    &		$<$\,0.8\,(3$\sigma$)		& km\,s$^{-1}$\\
\hspace{0.4cm}Mass of KOI-12.01  	   & $M_\mathrm{p}$	&	$<$\,10\,(3$\sigma$)  	& $M_\mathrm{Jup}$	\\
\hline
\noalign{\smallskip}
\textit{Stellar parameters} &								&																							 & \\
\hspace{0.4cm}Density			& $\rho_{\star}$ 						& 0.335  $\pm$ 0.080 													 &$\rho_{sun}$  \\
\hspace{0.4cm}Radius 	& $R_{\star}$ 							&1.63  $\pm$  0.15 														 &$R_{sun}$ \\
\hspace{0.4cm}Age			& $tau_{\star}$ 							&1.5 $\pm$ 0.5 																 &Gyr \\
\hspace{0.4cm}Mass		& $M_{\star}$ 							& 1.452  $\pm$  0.093 												 &$M_{sun}$ \\
\hspace{0.4cm}Distance		& $D$   								   &426  $\pm$  40 														   &pc \\
\hspace{0.4cm}Effective temperature    & $T_{\rm eff}$  					& 6820  $\pm$  120 														 &K \\
\hspace{0.4cm}Limb-darkening coefficients & $\epsilon_\mathrm{a}$ & 		0.334$\pm$0.007														 &  \\
							& $\epsilon_\mathrm{b}$ & 		0.124$\pm$0.013														 &  \\									
\hline
\hline
\noalign{\smallskip} 
Circular orbit     & & & \\
\hline
\noalign{\smallskip}
\textit{Orbital parameters} & & & \\
\hspace{0.4cm}Orbital period 		   & $P$ 					& 17.8552332$\pm$1.0$\times$10$^{-6}$				 & day \\
\hspace{0.4cm}Transit epoch 		   & $T_{0}$ 		        & 2454979.59601$\pm$0.5$\times$10$^{-4}$ 	             & BJD  \\
\hspace{0.4cm}Scaled semi-major axis & $a_\mathrm{p}/R_{\star}$  & 18.84$\pm$0.04  								 & \\
\hspace{0.4cm}Semi-major axis & $a_\mathrm{p}$  & 0.151$\pm$3$\times$10$^{-3}$  & au \\
\hspace{0.4cm}Orbital inclination 			& $i_\mathrm{p}$ 				& 88.90$\pm$0.02 				     & deg\\
\hspace{0.4cm}Impact parameter                  & $b$ 				&   0.362$\pm$7$\times$10$^{-3}$        &  \\
\hspace{0.4cm}Planet-to-star radii ratio 	& $R_\mathrm{p}/R_{\star}$ 	& 0.09049$\pm$8$\times$10$^{-5}$ 				   & \\
\hspace{0.4cm}Planet radius 	& $R_\mathrm{p}$ 	&	1.44$\pm$0.13	 & R$_{Jup}$\\
\hspace{0.4cm}Transit duration 					  & $t_\mathrm{d}$ 		&	  6.7827$\pm$2.6$\times$10$^{-3}$                          & hours \\
\hspace{0.4cm}Stellar reflex velocity   & $K$ 				& $<$\,0.51\,(3$\sigma$)              	& km\,s$^{-1}$\\
\hspace{0.4cm}Mass of KOI-12.01  	   & $M_\mathrm{p}$	& $<$\,8.7\,(3$\sigma$)  				 	& $M_\mathrm{Jup}$	\\
\hline
\noalign{\smallskip}
\textit{Stellar parameters}  & & & \\
\hspace{0.4cm}Limb-darkening coefficients & $\epsilon_\mathrm{a}$ & 0.335$\pm$0.008													 &  \\
							& $\epsilon_\mathrm{b}$ & 0.123$\pm$0.014														 &  \\	
\hline
\multicolumn{4}{l}{Notes: All parameters are derived from the analysis of the transit light curve (Sect.~\ref{photom}), except for the limits}\\
\multicolumn{4}{l}{on the Keplerian semi-amplitude and the mass of KOI-12.01 derived from the radial velocity analysis (Sect.~\ref{kep_fit}).}\\
\multicolumn{4}{l}{Notes: The derived Kepler transit epochs are given in Barycentric Dynamical Time.}\\
\end{tabular}
\label{table:tab_paramsfit}
\end{table*}

\section{Radial velocity analysis}
\label{RV_section}

\subsection{Keplerian fit}
\label{kep_fit} 

\begin{figure*}
\centering
\begin{minipage}[b]{0.9\textwidth}	  
\includegraphics[trim=1.5cm 1cm 0.8cm 2cm, clip=true,width=\columnwidth]{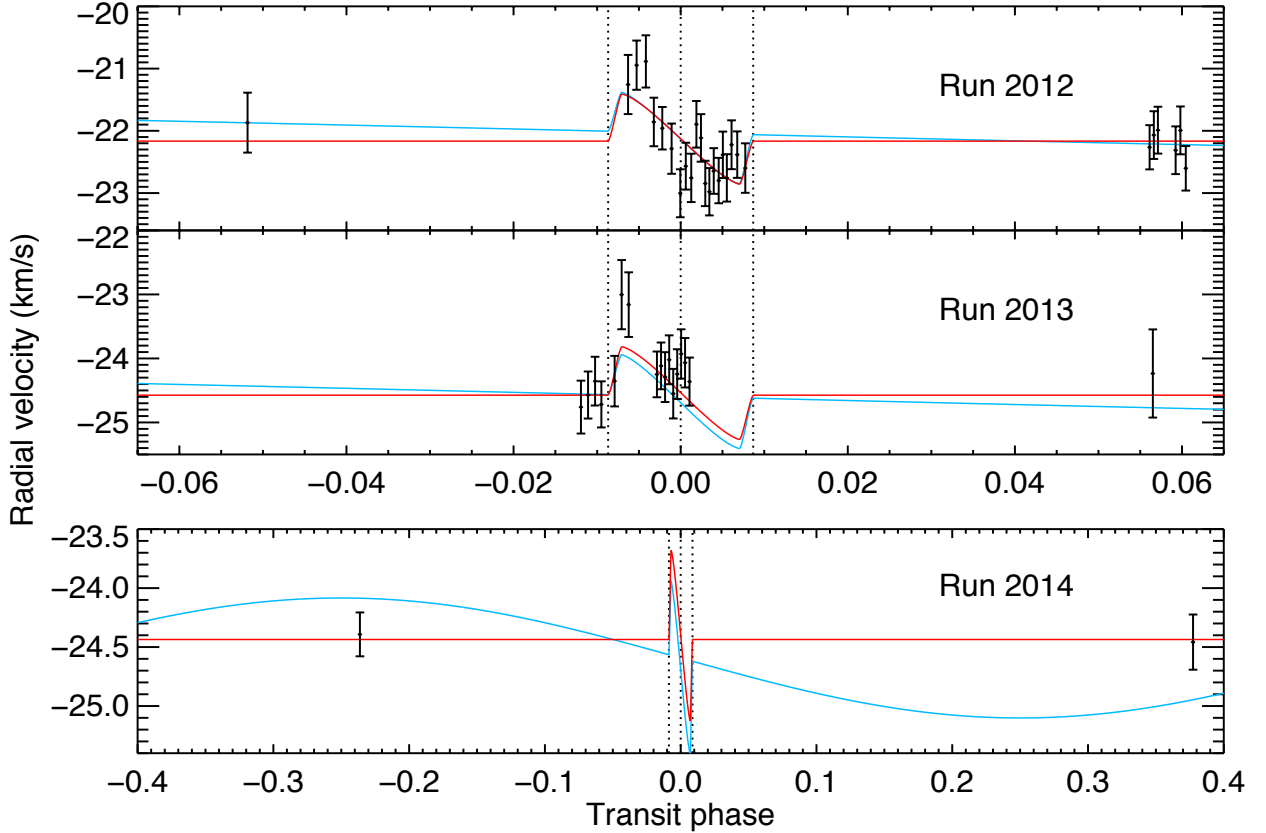}		
\end{minipage}	
\caption[]{Radial velocity measurements during Runs 2012, 2013 and 2014 (black points, from top to bottom), and their circular Keplerian + RM fit for a null semi-amplitude (red line) and its 3$\sigma$ upper limit (blue line). The Keplerian fit was performed on all measurements outside of the transit, and yields a total dispersion on the residuals of 0.21 and 0.28\,km\,s$^{-1}$, respectively. Variations in the asymmetric continuum of the CCFs (Fig.~\ref{fig:CCFs}) may cause the apparent differences in systemic velocities between the three runs. The RM model was adjusted to dataset 2012 only, but is consistent with dataset 2013. The two anomalous measurements in Run 2013 are the most redshifted RV values near ingress (vertical dashed lines show the times of mid-transit, first, and fourth contacts).} 
\label{fig:fit_kepler}
\end{figure*}

We fitted radial velocities outside of the transit in Runs 2012, 2013 and 2014 simultaneously, using a Keplerian orbit. The orbital period $P$ and transit epoch $T_{0}$ were fixed to their photometric values (Table~\ref{table:tab_paramsfit}). We computed the $\chi^2$ of the fit on a grid scanning a large range of values for the semi-amplitude $K$ of the radial velocity variations, and the systemic radial velocities for each dataset $\gamma_\mathrm{2012}$, $\gamma_\mathrm{2013}$, $\gamma_\mathrm{2014}$. We adjusted independently the systemic velocities because of possible variations in the asymmetric continuum of the CCFs, or residual instrumental drifts, between the three runs (Sect.~\ref{spec_obs}). Once the minimum $\chi^2$ and corresponding best values for these parameters were obtained, we calculated their error bars from an analysis of the $\chi^2$ variation (see e.g. \citealt{hebrard2002}).\\
Radial velocity measurements and their Keplerian model for a circular orbit are plotted in Fig.~\ref{fig:fit_kepler}. We found systemic radial velocities of -22.17$\pm0.09$, -24.57$\pm0.10$ and -24.44$\pm0.09$\,km\,s$^{-1}$ for Runs 2012, 2013 and 2014 respectively. By comparison the RV measured in 2011 are in the order of -18.5\,km\,s$^{-1}$ (\citealt{Santerne2012}). The significant variations in systemic velocity between the runs may be caused by variations in the asymmetric continuum of the CCFs, residual instrumental drifts, or the presence of a distant massive companion to KOI-12. However the small number of measurements over long timescales, the lack of RV standard stars in 2011 and possible systematics in SOPHIE HE mode (Sect.~\ref{spec_obs}) prevent us from concluding. The effects of the asymmetric continuum were however taken into account in the tomography analysis (Sect.\ref{results}), allowing us to refine the value for the systemic velocity of KOI-12 in Run 2012. The precision of our RV measurements and the sampling of the orbit are insufficient for a significant detection of the star's reflex motion, and we derived a $3\sigma$ upper limit of 0.51\,km\,s$^{-1}$ on the Keplerian semi-amplitude $K$ (Table~\ref{table:tab_paramsfit}). This parameter is mainly constrained by measurements in Run 2014, which are closer to quadrature-phase than measurements in Runs 2012 and 2013 taken near the transit epoch. Using the orbital inclination $i_\mathrm{p}$ and stellar mass $M_{\star}$ obtained in Sect.~\ref{photom}, together with their uncertainties, the limit for $K$ corresponds to a 3$\sigma$ upper mass limit $M_\mathrm{p}$=8.7\,$M_\mathrm{Jup}$ for KOI-12.01. This value is more stringent than the limit of 26.7\,$M_\mathrm{Jup}$ obtained by \citet{Santerne2012}, as the more recent datasets we used in Sect.~\ref{spec_obs} are of better quality with more measurements. It is also consistent with the maximum projected mass of 25.2\,$M_\mathrm{Jup}$ determined by \citet{Lillobox2015}. We performed the same analysis assuming an eccentric orbit, setting the eccentricity to its 3$\sigma$ upper limit of 0.72. We found that the 3$\sigma$ upper limits on $K$ and $M_\mathrm{p}$ increased to 0.8\,km\,s$^{-1}$ and 10\,$M_\mathrm{Jup}$, which is taken hereafter as the final upper limit on the mass of KOI-12.01.

\subsection{Rossiter-McLaughlin anomaly}
\label{rm_ano} 

\begin{figure}[]   
\includegraphics[trim=1cm 10cm 2cm 0cm, clip=true,width=\columnwidth]{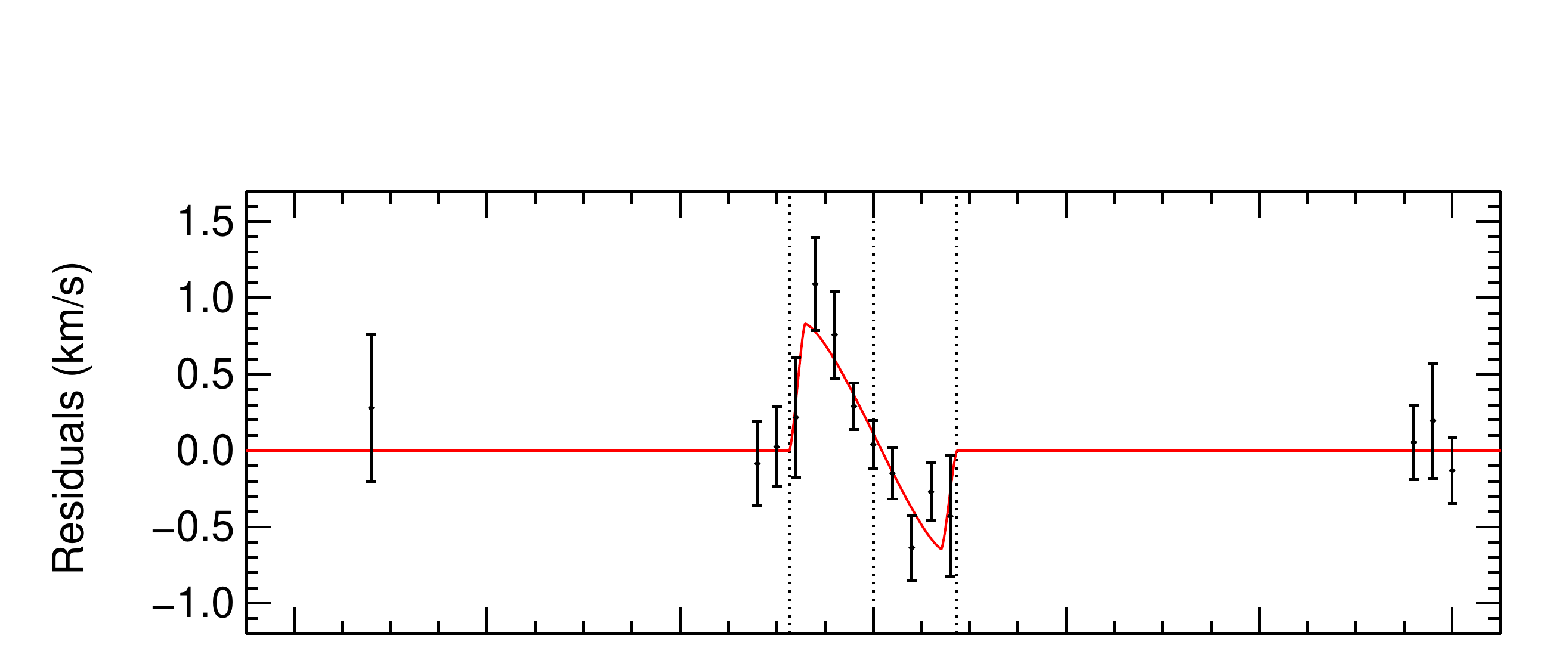}
\includegraphics[trim=1cm 2cm  2cm 11.5cm, clip=true,width=\columnwidth]{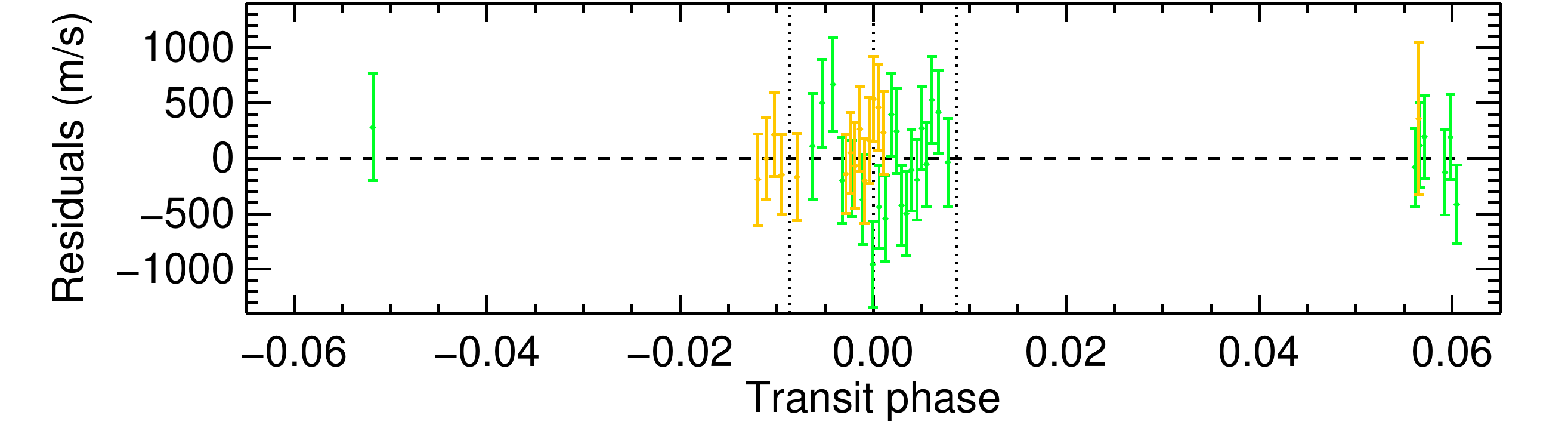}
\caption[]{\textit{Upper panel:} Error-weighted average of Runs 2012 and 2013 RV measurements (black points), following the subtraction of the systemic velocities and assuming no reflex stellar motion ($K$=0\,km\,s$^{-1}$). The RM adjustment with $\lambda$=16.0$\stackrel{+9.9}{_{-10.4}}^\circ$ and $v$sin$i_{\star}$=123.5$\pm$22.5\,km\,s$^{-1}$ is overplotted with a red line. Keplerian residuals are first calculated in each run and then grouped in common phase bins. We excluded the two anomalous RV measurements in dataset 2013 (Fig.~\ref{fig:fit_kepler}). \textit{Lower panel:} Residuals from the Keplerian + RM fit, without binning. Green points correspond to Run 2012, orange points to Run 2013.}
\label{fig:residus_RM}
\end{figure}

Spectra in Run 2013 were acquired in poor weather conditions, and their tomographic analysis (see Sect.~\ref{check_runB}) allowed systematic effects to be identified in two of the radial velocity measurements (see Fig.~\ref{fig:fit_kepler}). In addition, most of Run 2013 measurements are grouped near the transit center, which makes it difficult to constrain the shape of the RM anomaly. In contrast, Run 2012 has a good sampling of the whole transit with data of good quality. We thus applied the analytical approach developed by \citet{Ohta2005} to model the form of the RM anomaly in this run only. The semi-amplitude was fixed to 0\,km\,s$^{-1}$ and the systemic velocity was fixed to its best-fit value derived in Sect.~\ref{kep_fit} for a circular orbit. The stellar limb-darkening coefficients, the transit parameters $R_\mathrm{p}/R_{\star}$, $a_\mathrm{p}/R_{\star}$ and $i_\mathrm{p}$, $P$ and $T_{0}$ were fixed to their photometric values (Table~\ref{table:tab_paramsfit}). The best-fit values and corresponding error bars for the sky-projected obliquity $\lambda$ and the projected stellar rotation velocity $v$sin$i_{\star}$ were determined in the same way as in Sect.~\ref{kep_fit} using $\chi^2$ analysis. \\
The best model for the RM anomaly is shown in Fig.~\ref{fig:fit_kepler}. It was obtained with $v$sin$i_{\star}$=117$\pm$25\,km\,s$^{-1}$, which is 2\,$\sigma$ higher than the value derived from the spectral analysis in Sect.~\ref{spectralana}. The analytical formulae developed by \citet{Ohta2005} do not provide a good estimate of the projected stellar rotation velocity for rapidly rotating stars like KOI-12 (\citealt{Hirano2010}). Yet, we also analyzed the RM anomaly using the \citet{gimenez2006} equations and found a similar value with $v$sin$i_{\star}$=107$\pm$23\,km\,s$^{-1}$. Both methods do not take into account the effects of macroturbulence, which are expected to be important for KOI-12 (Sect.\ref{spectralana}). This may have biased the estimates of the projected rotational velocity toward higher values, although a macroturbulence in the order of 18\,km\,s$^{-1}$ cannot explain an overestimation by about 50\,km\,s$^{-1}$. The value for $v$sin$i_{\star}$ will be determined more accurately using the technique of Doppler tomography in Sect.~\ref{doppler}. Nonetheless, analysis of the RM anomaly using the analytical formulae of \citet{Ohta2005} and \citet{gimenez2006} provides a useful first-order estimate of the obliquity value, and we measured $\lambda$ = 5.8$\stackrel{+13.1}{_{-15.5}}^\circ$ using both methods. The fitted obliquity remained consistent within its uncertainties when we increased to their 3$\sigma$ upper limits the Keplerian semi-amplitude $K$ ($\lambda$ = -14.7$\stackrel{+14.4}{_{-15.8}}$\,$^\circ$) or the eccentricity $e$ ($\lambda$ = 1.7$\stackrel{+13.9}{_{-16.1}}$\,$^\circ$), as in both cases the Keplerian velocity variations remain small on the transit duration timescale (see Fig.~\ref{fig:fit_kepler}). Finally, we checked the consistency of Run 2013 with the RM anomaly detected in Run 2012 by fitting them simultaneously, assuming no stellar reflex motion. The best RM model is plotted in Fig.~\ref{fig:residus_RM} and corresponds to an obliquity of 16.0$\stackrel{+9.9}{_{-10.4}}^\circ$ and a projected rotational velocity of 123.5$\pm$22.5\,km\,s$^{-1}$, in agreement with the values obtained using Run 2012 only. The two RV measurements affected by systematic effects in dataset 2013 were excluded from the simultaneous fit, as they are strongly redshifted and force the stellar rotational velocity to a higher value (see the tomographic analysis and Fig.~\ref{fig:res_2013}). Nonetheless, we found that including them in the fit does not change significantly the derived obliquity.\\ We thus conclude to the detection of the RM anomaly of KOI-12.01, which shows a prograde orbit consistent with an aligned system in both transit observations. We note, though, that the large uncertainties derived for the obliquity do not rule out a small misalignment.\\

\section{Doppler Tomography}
\label{doppler} 

\subsection{Parameter fitting method}
\label{model} 

\begin{figure}
\centering
\includegraphics[trim=1.5cm 12cm 3cm 0cm, angle=180,clip=true,width=\columnwidth]{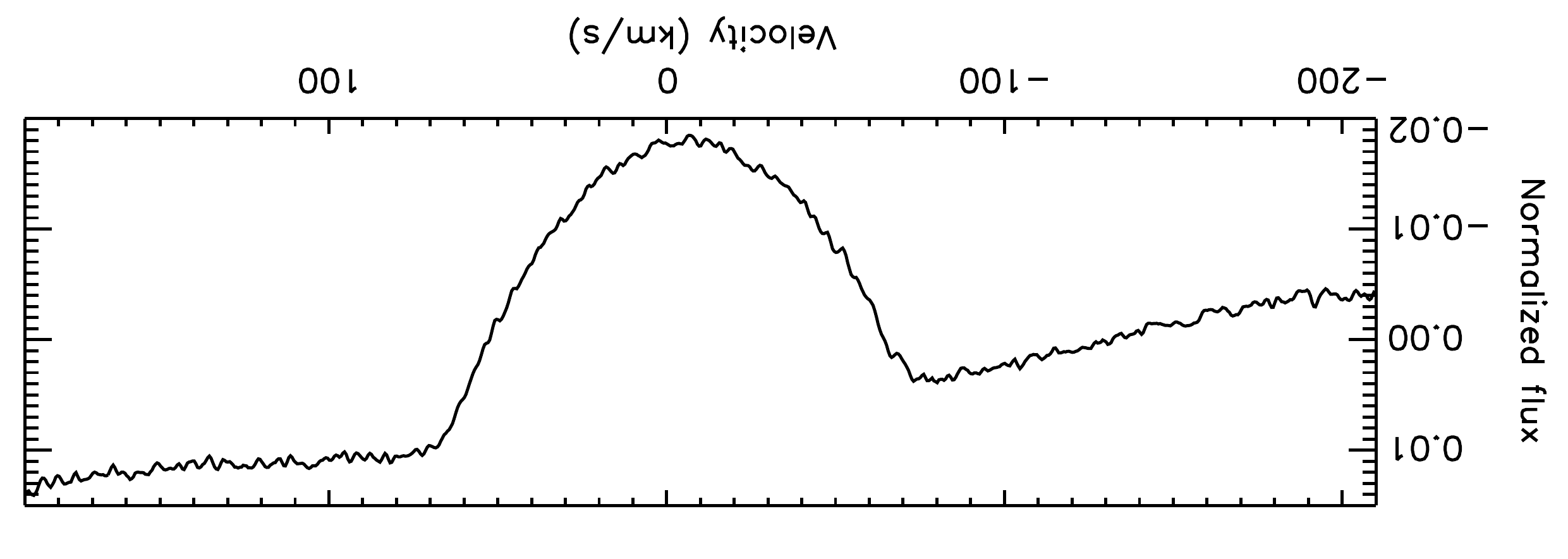}	
\caption[]{Mean of the CCFs observed outside of the transit in Run 2012, as a function of radial velocities relative to the star. Before calculating the average, the CCFs in each run were shifted from the velocity frame of the Solar system barycenter where they are computed, to the frame of KOI-12 using the best-fit systemic velocities derived in Sect.~\ref{kep_fit}. Note the strong tilt of the asymmetric continuum.}
\label{fig:CCFs}
\end{figure}

We then used Doppler tomography to obtain complementary values for the system properties and investigate their consistency with the fit to the RM anomaly. Although this latter analysis showed a prograde orbit and excluded a high misalignment, it could not indeed place strong constraints on the obliquity and yielded an anomalously high value for the projected stellar rotation velocity. We expect a more reliable value for this parameter to be derived using tomography, as this method is well suited to fast rotators such as KOI-12 (\citealt{cameron2010}) and takes into account the effects of macroturbulence through the direct modeling of the local line profile. We used prior constraints from photometry to increase the accuracy of our results. For the reasons given in Sect.~\ref{rm_ano} we based our analysis on the spectra secured during Run 2012, and in a second time we checked the consistency of dataset 2013 with our results (Sect.~\ref{check_runB}). The CCFs obtained with the SOPHIE pipeline were fitted using the technique developped by \citet{cameron2010a}, which relies on the decomposition of the CCF into its different components: the stellar line profile model outside the transit is a limb-darkened rotation profile convolved with a Gaussian corresponding to the intrinsic photospheric line profile plus instrumental broadening. The planet occultation is modeled as a Gaussian 'bump' in the stellar line profile, whose spectral location depends on the planet position in front of the stellar disk during the transit. The continuum of KOI-12 CCFs is asymmetric and strongly tilted (Fig.~\ref{fig:CCFs}). While stellar activity, moonlight and telluric contamination, and random alignments between photospheric lines and mask lines at arbitrary RV shifts are known to create anomalies in the profile of the CCF (the so-called sidelobe patterns), here the observed tilt may be caused by an imperfect removal of the instrument blaze function. While this pattern may decrease the quality of the CCFs adjustment, we noted that it remains fixed over a few consecutive nights and can thus be corrected for by subtracting the average of the difference between each CCF and its model line profile (\citealt{cameron2010a}). To prevent removing inadvertently KOI-12.01 signature, however, only observations outside the transit were used to calculate this correction. \\

The tomographic model depends on the same parameters as the Keplerian + RM model (Sect.~\ref{RV_section}), with the addition of the local line profile width $s$=FWHM/(2\,$\sqrt{2\,ln\,2}$\,$v$sin$i_{\star}$), expressed in units of the projected stellar rotational velocity (\citealt{cameron2010a}). FWHM is the full-width at half-maximum of the non-rotating stellar line profile convolved with the instrument profile\footnote{The FWHM$_{CCF}$ introduced in Sect.~\ref{spec_obs} is instead the full-width at half-maximum of the rotationally-broadened CCF}. The effects of macroturbulent velocity on the width of the local line are thus already accounted for by the FWHM, independently of $v$sin$i_{\star}$. We fixed the period and the mid-transit time to their values determined with a high precision using the Kepler photometry (Sect.~\ref{photom}). The semi-amplitude was set to 0\,km\,s$^{-1}$ (Sect.~\ref{kep_fit}) and we used a linear limb-darkening law with coefficient $\epsilon$. We discuss the influence of these parameters values and assumptions on the results in Sect.~\ref{check_mod}. All other parameters were adjusted using the following merit function:
\begin{equation}
\begin{split}
\chi^2=&\sum\limits_{i}^{n_{CCF}} \sum\limits_{j}^{n_v} \left[ \frac{f_{i,j}(obs)-f_{i,j}(mod)}{\sigma_{i}}  \right]^2     + \\
&\sum\limits_{a_\mathrm{p}/R_{\star}, i_\mathrm{p}, R_\mathrm{p}/R_{\star}}\left[ \frac{x_{tomo}-x_{photo}}{\sigma_{x_{photo}}}  \right]^2 ,
\end{split}	
\label{eq:stdev}  
\end{equation}
where $f_{i,j}$ is the flux at velocity point $j$ in the $i$th observed or model CCFs. The error on the flux, $\sigma_{i}$, is supposed constant for a given CCF. Photometry cumulated over seventy-four Kepler transits is expected to provide a more precise determination for the radius, semi-major axis and orbital inclination of KOI-12.01 than tomography. The model value of these parameters $x_{tomo}$ is constrained directly through the $\chi^2$ using the best-fit value $x_{photo}\pm\sigma_{x_{photo}}$ from the transit light-curve analysis (Table~\ref{table:tab_paramsfit}). The fit to the data is performed in three steps:
\begin{enumerate}
\item To identify all possible local minima in the parameters space, we analyze $\chi^2$ variations. A given parameter is pegged at various trial values, and for each trial value we run a Levenberg-Marquardt minimization algorithm, allowing all other parameters to vary freely.
\item We use the residuals between the CCFs and their best-fit model profile from step 1 to estimate the noise on the CCFs. The procedure is described in details in Sect.~\ref{noise}. Note that the value of $\sigma_{i}$ has little influence on the results of step 1.
\item We refine the best-fit parameters values and calculate their uncertainties using MCMC. The Markov chain is constructed by repeatedly computing the merit function for a sequence of parameter values. At each step of the chain, the difference $\Delta \chi^2$ is calculated and the new set of parameters is accepted if $\Delta \chi^2 <$0, or if $\Delta \chi^2 >$0 with a probability equal to the ratio of likelihoods. The next set of parameters is computed by applying Gaussian perturbations to the preceding set: $x_{i+1}=x_{i}+Gauss(0,\sigma_{x})$. The standard deviation $\sigma_{x}$ is specific to each parameter, and is reevaluated for all parameters with a constant frequency, so that a change $x_{i+1}=x_{i}+\sigma_{x}$ would correspond to $\Delta \chi^2=1$. This ensures that the space of each parameter is correctly sampled and that their posterior probability distributions have the correct variances. We start several independent chains from different sets of parameters chosen randomly near the local minima found in step 1. This limits the burn-in phase to only a few hundred points, after which the chain is in a steady state and all accepted steps are used to compute the probability distributions. Although the chains are let free to explore all values for the orbital inclination and obliquity, we use the following conventions to compute the final distributions: 0$^\circ\,\le\,i_{p}\,\le\,$90$^\circ$ ; -180$^\circ\,\le\,\lambda\,\le\,$180$^\circ$. Accordingly, all couples ($i_{p}>$90$^\circ,\lambda$) are tranformed into ($ $180$^\circ-i_{p},-\lambda $) as the two possibilities are equivalent with respect to the tomography model equations. The medians of the distributions are chosen as the final best-fit values for the model parameters. Their 1$\sigma$ uncertainties are obtained by finding the intervals on both sides of the median that contain 34.15\% of the accepted steps. 
\end{enumerate}

\subsection{Determination of CCFs errors}
\label{noise} 

\begin{figure}[]   
\includegraphics[trim=1cm 2cm 2.5cm 2.5cm, clip=true,width=\columnwidth]{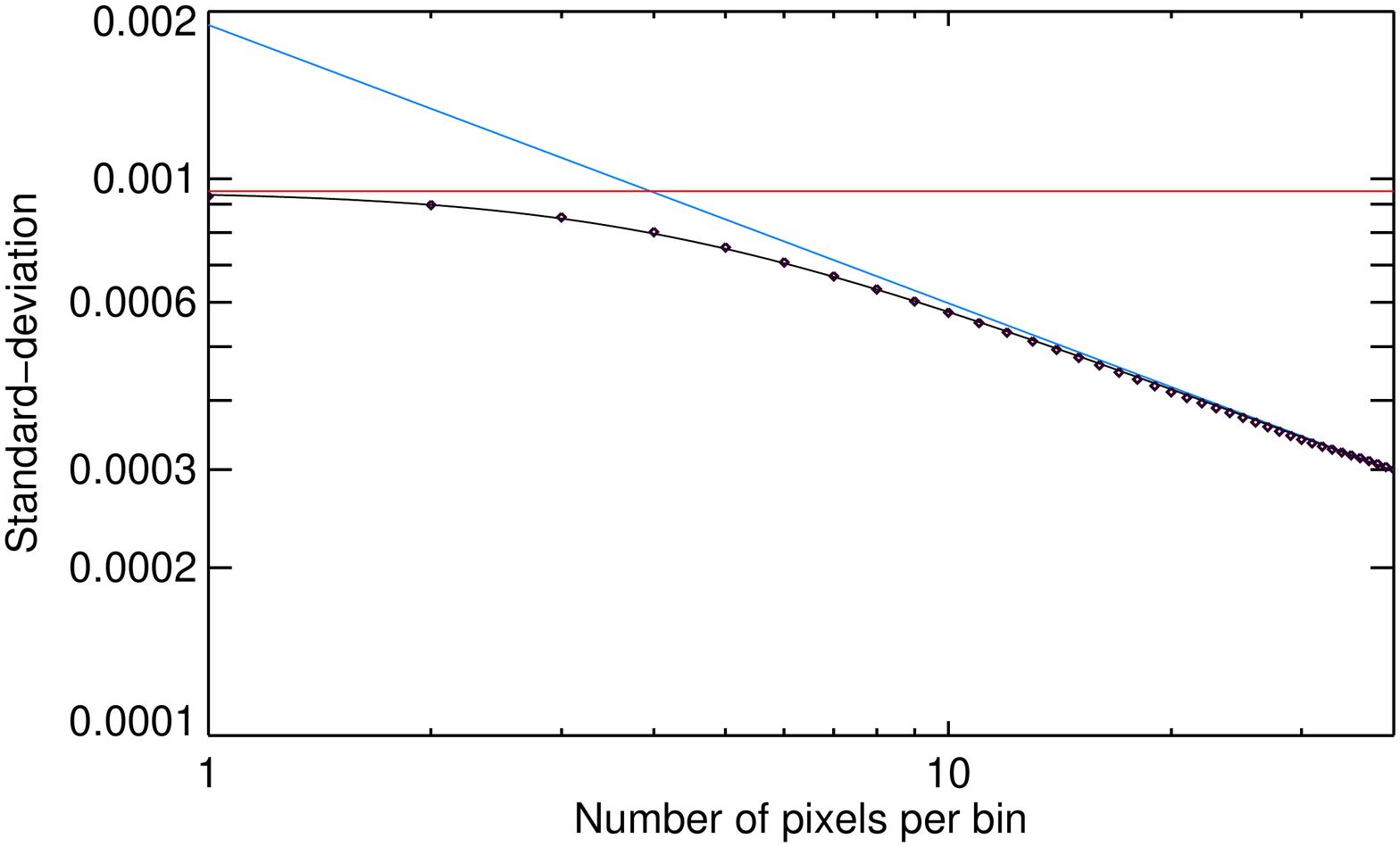}
\caption[]{Standard deviation of velocity-binned residuals between a CCF and its best-fit model, as a function of the binning factor (black diamonds). A single pixel is 0.5\,km\,s$^{-1}$ wide. The best fit (black line) is a harmonic combination of an uncorrelated noise component, inversely proportional to the square root of the binning factor (blue line), and a constant correlated noise component (red line).}
\label{fig:fit_variance}
\end{figure}

Because errors on the CCF data points are unknown, we attributed at first to each pixel of a given CCF a constant error equal to the dispersion of the residuals between this CCF and its best-fit model profile. However the CCFs were calculated with the SOPHIE pipeline at a velocity resolution of 0.5\,km\,s$^{-1}$, while the spectra were observed in the HE mode at an instrumental resolution of about 7.5\,km\,s$^{-1}$. The residuals were thus found to be strongly correlated, which can lead to underestimation of error bars on the derived parameters. To derive reliable uncertainties on the model parameters using $\chi^2$ statistics, errors on the CCF pixels must be independent and follow a Gaussian random distribution (a 'white' noise). We thus decided to retrieve directly the non-correlated gaussian component of the CCFs noise using an analysis of the residuals variance inspired from \citet{Pont2006}:
\begin{enumerate}
	\item We calculate the CCF residuals defined as the difference between a CCF and its best-fit model obtained in step 1 of the fitting procedure (Sect.~\ref{model}). 
	\item We average the residuals within an interval of length $n_{bin}$ pixels (the binning factor). This interval slides along the entire residual range, and we record the residual means for every possible position where the interval contains a different set of $n_{bin}$ successive pixels.
	\item To each value of the binning factor corresponds a distribution of residuals means, for which we calculate the standard deviation $\sigma(n_{bin})$. This parameter represents the characteristic dispersion of the CCF pixels binned at a given resolution. This dispersion is the result of the combination of the intrinsic Gaussian noise in individual measurements and the correlation between them. \\
\end{enumerate}
After calculating the standard deviation for every CCF, we found empirically that it is always well represented by a quartic harmonic combination of a white and a red noise components:
\begin{equation}
\sigma^2(n_{bin})=\left(\left(\frac{n_{bin}}{\sigma_{Uncorr}^2}\right)^{2}+\left(\frac{1}{\sigma_{Corr}^2}\right)^{2}    \right)^{-\frac{1}{2}}.  
\label{eq:stdev}   			
\end{equation}
where $\sigma_{Uncorr}/\sqrt{n_{bin}}$ can be understood as the result of the intrinsic uncorrelated noise after the binning of $n_{bin}$ pixels, and $\sigma_{Corr}$ is a constant term characterizing the correlation between the binned pixels. A typical curve for the standard deviation as a function of the binning factor is shown in Fig~\ref{fig:fit_variance}. To validate Eq.~\ref{eq:stdev} we created thousands of synthetic CCF residual tables in which each point is drawn from a Gaussian random distribution. These residuals were interpolated onto a velocity table with higher resolution to make them equivalent to the CCF residuals resulting from the model fits to the observed CCFs, by creating correlation between adjacent pixels as observed in the data. For each synthetic table we then calculated $\sigma(n_{bin})$ using the above method, and verified that it was always best described by Eq.~\ref{eq:stdev}.  \\
As can be seen in Fig~\ref{fig:fit_variance} the uncorrelated noise in the CCFs becomes dominant for binning factors higher than about 15, because the velocity width of the bins is then larger than SOPHIE instrumental resolution. In previous tomography studies, the issue of correlated noise was circumvented by binning CCFs at the instrumental resolution (e.g. \citealt{miller2010}). To avoid possible loss in resolution, we did not bin the data and attributed to all pixels in a given CCF the same uncorrelated noise $\sigma_{Uncorr}$. This value represents the theoretical intrinsic Gaussian noise (the blue line in Fig~\ref{fig:fit_variance}) of an unbinned pixel. With CCFs errors now corresponding to uncorrelated noise only, uncertainties on the model parameters derived from the fit of the CCFs can be correctly estimated using $\chi^2$ statistics, independently of the CCF sampling.\\

\subsection{Results}
\label{results}

\begin{figure*} 
\centering
\begin{minipage}[b]{0.5\textwidth}	
\includegraphics[trim=0cm 8.8cm 1.8cm 1cm,clip=true,width=\columnwidth]{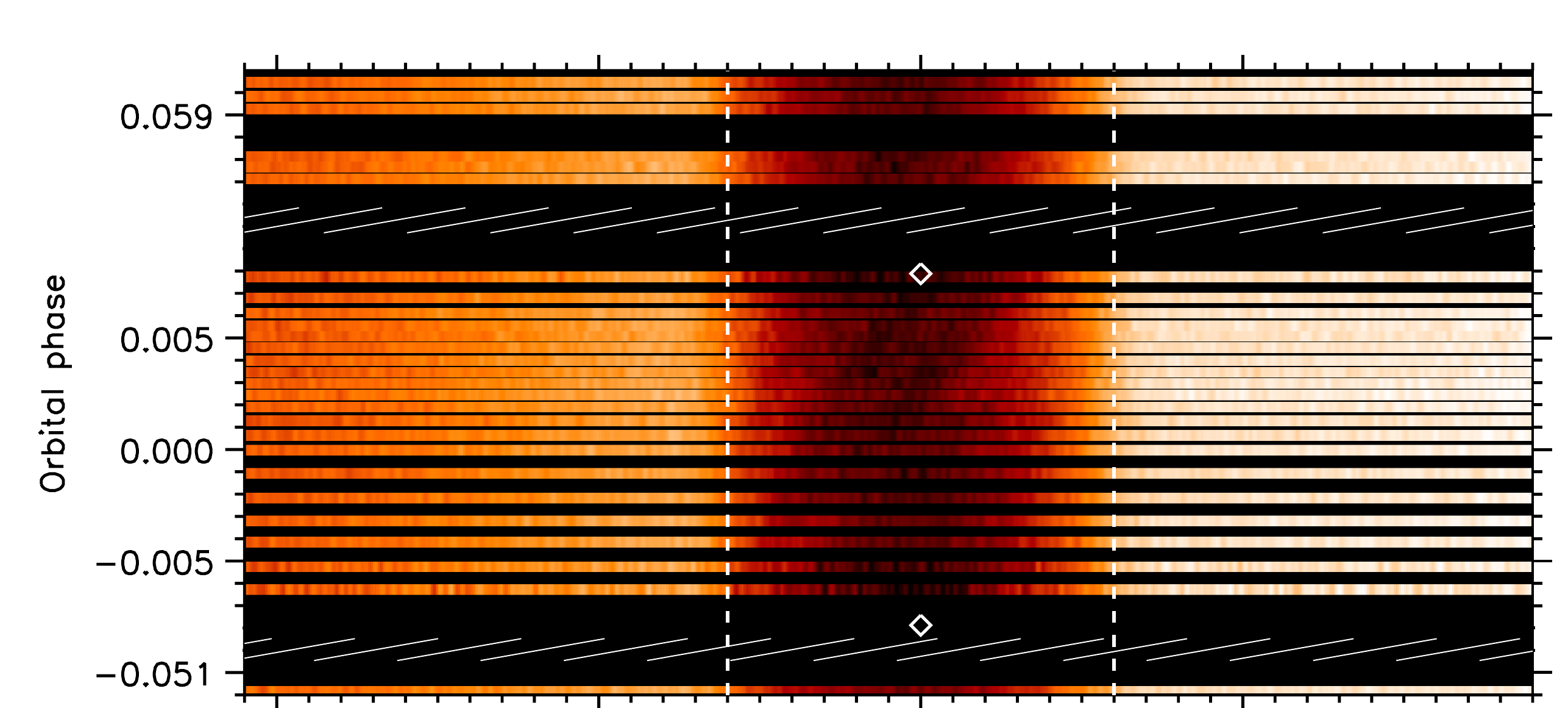}	
\includegraphics[trim=0cm 4.cm 1.8cm 2cm,clip=true,width=\columnwidth]{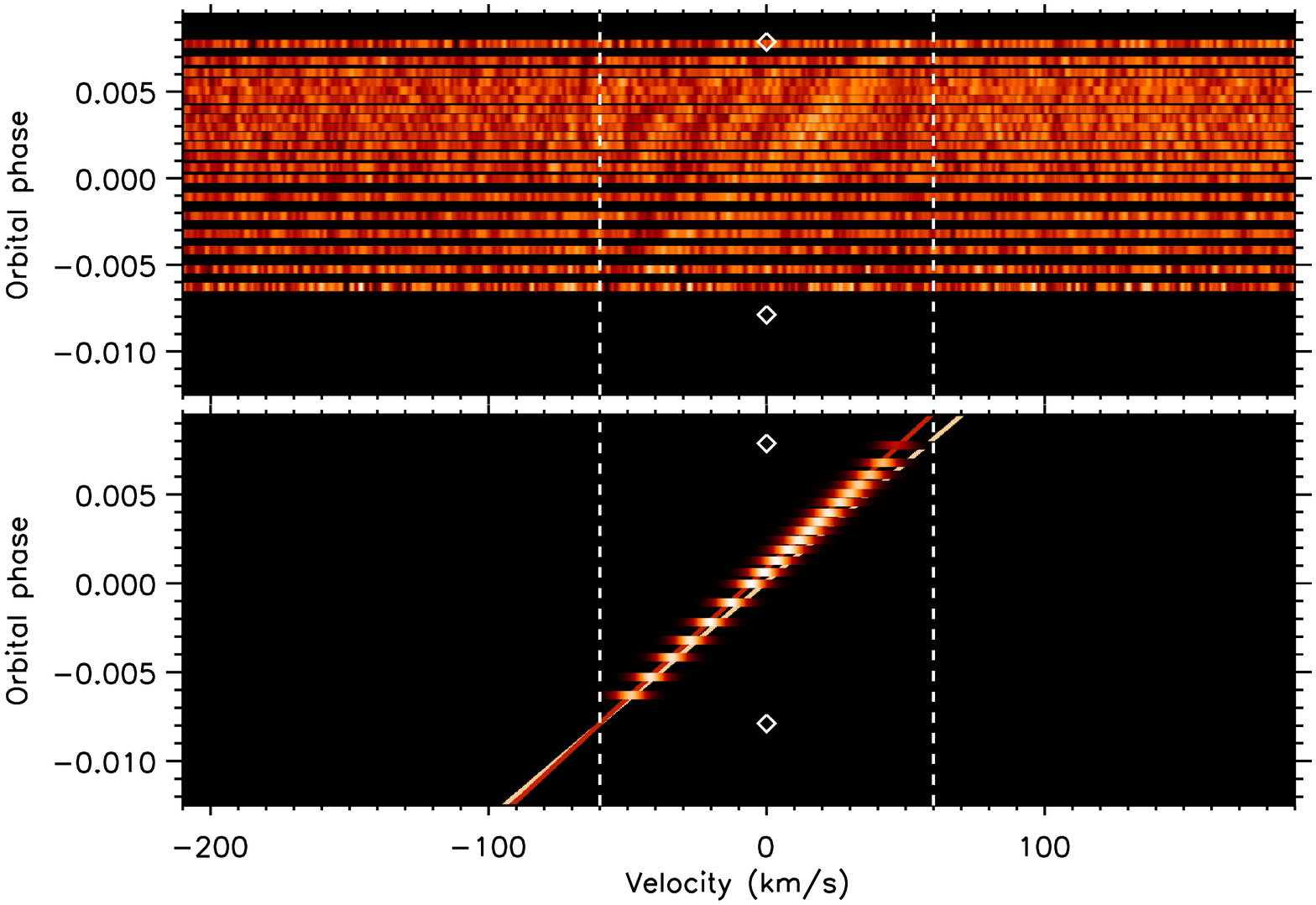}	
\includegraphics[trim=0cm 2cm 1.8cm 9.7cm,clip=true,width=\columnwidth]{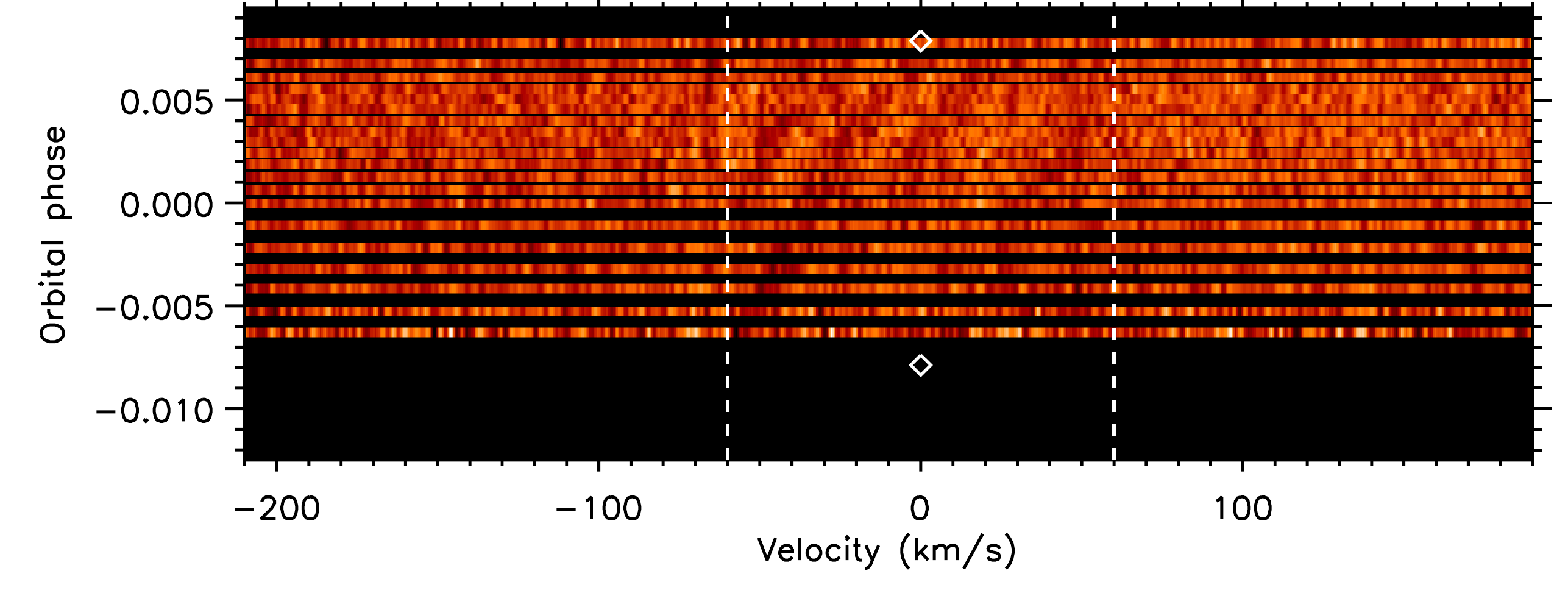}
\end{minipage}
\caption[]{Maps of the time-series CCFs from Run 2012, as a function of radial velocity relative to the star (in abscissa) and orbital phase (in ordinate, increasing vertically). Flux values increase from dark red to white. Vertical dashed white lines are plotted at $\pm$ $v$sin$i_{\star}$, and white diamonds indicate the time of the 1$^{st}$ and 4$^{th}$ contacts. \textit{Top panel}: CCFs produced by the SOPHIE pipeline. Dashed white areas separate the different datasets taken over three successive nights (see Sect.~\ref{spec_obs}). The tilt of the continuum is visible as a flux increase from negative to positive velocities (see also Fig.~\ref{fig:CCFs}), and does not vary significantly from one night to the other. \textit{Upper middle panel}: Map of the transit residuals after subtracting the model stellar profile. The signature of KOI-12.01 is the bright and wide feature that crosses the entire width of the line profile from ingress to egress. \textit{Lower middle panel}: Best-fit model for KOI-12.01 transiting signature, emphasized by the dark brown line (extended beyond the first/fourth contact and $\pm$ $v$sin$i_{\star}$ for the sake of clarity) obtained with $\lambda$ = 12.5$\stackrel{+3.0}{_{-2.9}}$. This is consistent with the asymmetry of the signature visible in the above panel, whereas an aligned orbit would correspond to the lighter brown line. \textit{Bottom panel}: Overall residual map after the further subtraction of the model planet signature. The faint, alternating bright and dark parallel streaks may be caused by stellar intrinsic variability.}
\label{fig:residus_final}
\end{figure*}

In this section we present our results from the tomographic analysis of Run 2012 time-series CCFs, which are plotted in the top panel of Fig.~\ref{fig:residus_final}. Much information is readily available in the residual maps obtained by subtraction of the out-of transit model stellar line profile (Fig.~\ref{fig:residus_final}, upper middle panel). The transit is revealed as a bright streak confined between $\pm$ $v$sin$i_{\star}$, and whose spectral positions over time correspond to the radial velocity of the stellar surface regions occulted by KOI-12.01. For comparison, Fig.~\ref{fig:residus_final} (lower middle panel) shows the signature of a transiting planet that appears as a straight line in the residual maps. Indeed, its velocity can be written as $v_{p}=vsini_{\star}\,x_{\bot}$, where $x_{\bot}$ is the distance between the region occulted at orbital phase $\phi$ and the stellar spin axis,
\begin{equation}
x_{\bot}=a_{p}/R_{\star}\,(\,cos(\lambda)\,\sin(2\,\pi\,\phi)\,-\,\sin(\lambda)\,\cos(i_{p})\,\cos(2\,\pi\,\phi)\,). 
\label{eq_distaxe}
\end{equation}
During the transit, $\phi\sim0$ and the velocity of the signature approximates to a linear function of the orbital phase
\begin{equation}
v_{p}=vsini_{\star}\,a_{p}/R_{\star}\,(\,\cos(\lambda)\,2\,\pi\,\phi\,-\,\sin(\lambda)\,\cos(i_{p})\,). 
\label{eq_linvit}
\end{equation}
The signature of KOI-12.01 travels from negative to positive velocities during the transit, which shows unambiguously that it is on a prograde orbit, occulting first the blueshifted and then the redshifted regions of the stellar disk. As can be seen in Fig.~\ref{fig:residus_final} (upper middle panel), the signature trajectory is roughly symmetric, ruling out a high misalignement consistently with the analysis of the RM anomaly in Sect.~\ref{rm_ano}. However there is a hint that ingress occurs at a slightly higher absolute velocity ($\sim$60\,km\,s$^{-1}$) than egress ($\sim$50\,km\,s$^{-1}$), the former being close to the best-fit projected stellar rotation velocity (Table~\ref{tab_tomo}). This would be consistent with a moderately inclined orbit, the transit beginning near the equatorial plane and ending at a higher latitude. Finally, faint features are visible in the overall residual maps after subtraction of the model planet signature (bottom panel in Fig.~\ref{fig:residus_final}). As in the case of WASP-33 (\citealt{cameron2010}), such features could be due to oscillations of the stellar surface which move alternatively inward or outward depending on the longitude. 
However, the prograde motion of these signatures and their linearity in the phase/velocity residual maps suggest the presence of active regions carried around by the rotation of the star.\\

\begin{figure*}
\centering
\begin{minipage}[b]{0.9\textwidth}		
\includegraphics[trim=4cm 2cm 7cm 1cm, clip=true,width=\columnwidth]{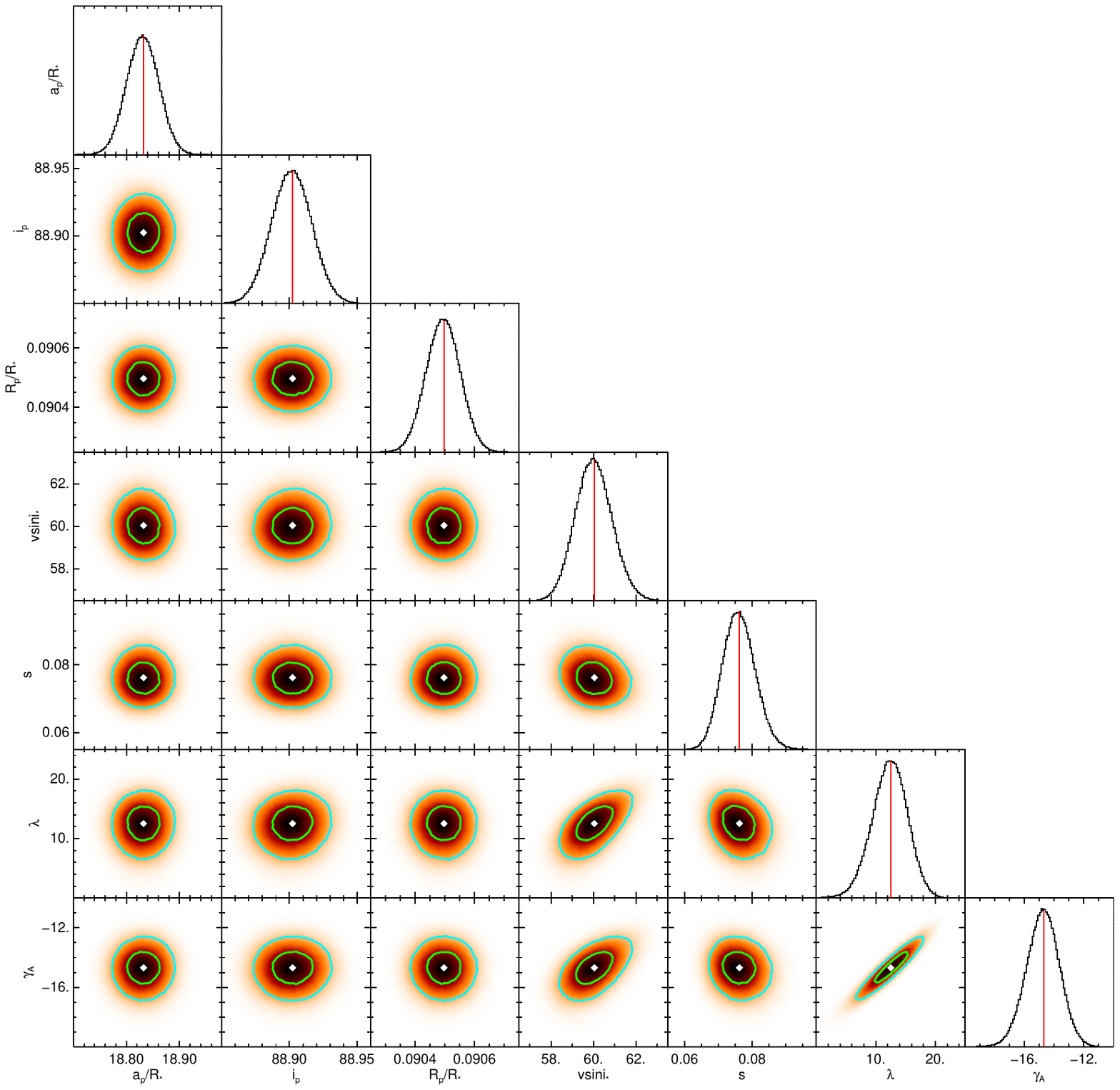}	
\end{minipage}
\caption[]{Correlation diagrams for the probability distributions of the tomographic model parameters. Priors from photometry were used to constrain $a_\mathrm{p}/R_{\star}$, $i_\mathrm{p}$ and $R_\mathrm{p}/R_{\star}$. The green and blue lines show the 1 and 2$\sigma$ simultaneous confidence regions that contain respectively 39.3\% and 86.5\% of the accepted steps. 1D histograms correspond to the distributions projected on the space of each line parameter. The red line and white point show median values. Units are the same as in Table~\ref{tab_tomo}.}
\label{fig:distrib_mcmc}
\end{figure*}

Only one set of parameters values was found to reproduce well the observations during step 1 of the fitting procedure (Sect.~\ref{model}). We ran eight MCMC chains of 10$^4$ accepted steps, with an average acceptance rate of 25\%. The posterior probability distributions for all model parameters are shown in Fig.~\ref{fig:distrib_mcmc}, along with the marginalized 1D distributions which are well represented by Gaussians and allowed us to derive tight constraints on the inferred best-fit values (Table.~\ref{tab_tomo}). The combined analysis and the use of priors on $a_\mathrm{p}/R_{\star}$, $R_\mathrm{p}/R_{\star}$, and $i_\mathrm{p}$ reduced the uncertainties on these parameters. We found $v$sin$i_{\star}$=60.0$\stackrel{+0.8}{_{-0.9}}$\,km\,s$^{-1}$, lower by about 2\,$\sigma$ than the value obtained using the Fourier transform method but in good agreement with the value derived from the \citealt{Boisse2010} relationship (Sect.~\ref{spectralana}). This confirms that KOI-12 is one of the most rapidly rotating star known to host a transiting planet so far. With such a large rotational broadening, the value derived for the local line width $s$=0.076$\pm$5$\times$10$^{-3}$ corresponds to a FWHM for the local stellar and instrumental profile of about 18\% of the projected rotation velocity (see the definition of $s$ in Sect.~\ref{model}).\\ 
The best-fit value for the obliquity $\lambda$=12.5$\stackrel{+3.0}{_{-2.9}}$ is consistent with the RM anomaly fit, and the visual analysis of the residual maps that hinted to an  asymmetry in the signature of KOI-12.01 (see Fig.~\ref{fig:residus_final}). The reduced uncertainties allow us to conclude to the low-obliquity of KOI-12.01 orbit.\\ 
The systemic velocity obtained from Gaussian-fitted RV measurements (Sect.~\ref{kep_fit}) is significantly different from the value derived from tomography. This is likely because the latter method is based on a direct modeling of the CCFs which takes into account the asymmetric non-flat continuum of the CCF (Fig.~\ref{fig:CCFs}), while a Gaussian adjustment with a flat continuum model introduces systematic shifts in the estimation of the velocity. Nonetheless this does not prevent to correctly sample the Keplerian velocity curve, as the patterns remain similar over a few nights and the velocity shifts are thus constant in a given dataset.\\

Finally, the diagrams in Fig.~\ref{fig:distrib_mcmc} show no parameter correlations except between $\lambda$, $\gamma_{2012}$ and $v$sin$i_{\star}$, and we therefore studied the influence of these three parameters on the model. The quality of the observed CCFs and the lower phase coverage during the first half of the transit allow for some indetermination in the spectral position of the planetary signature in the stellar reference frame. This velocity reference depends directly on the systemic velocity, and a higher value for $\gamma_\mathrm{2012}$ is equivalent in the residual maps to a shift of the planetary trace toward negative velocities relative to the star. This requires in turn that the planet occults the blue regions of the stellar disk for a larger amount of time, which can only be obtained with a higher obliquity since the impact parameter is tightly constrained by the photometry priors. There is thus a strong correlation between the systemic velocity and the obliquity, and an increase in these parameters also requires an increase in the rotational velocity since the shifted planetary signature must remain within the range of $\pm\,v$sin$i_{\star}$ during the transit.

\begin{table*}
\caption{Doppler tomography analysis, with and without constraints from the photometry analysis. Pertinent values from Table~\ref{table:tab_paramsfit} have been reported in the first column for a more direct comparison. The final values for the system properties are derived from the tomography + photometry analysis.}
\begin{tabular}{lccccl}
\hline
\noalign{\smallskip}  
\textbf{Parameter}   & \textbf{Symbol} 	& \textbf{Photometry}       & \textbf{Tomography}              & \textbf{Tomography + Photometry} 		   & \textbf{Unit} \\   			
\noalign{\smallskip}
\hline
\hline
\noalign{\smallskip}
Scaled semi-major axis 		& $a_\mathrm{p}/R_{\star}$& 18.84$\pm$0.04  		  &17.7$\stackrel{+1.4}{_{-0.7}}$     &\textbf{18.83$\pm$0.03}$^\dagger$   & \\
Orbital inclination  	 & $i_\mathrm{p}$ 		  & 88.90$\pm$0.02 		&89.3$\stackrel{+0.4}{_{-0.7}}$ 		&\textbf{88.902$\pm$0.015}$^\dagger$       & deg \\
Impact parameter &  $b$	&   0.362$\pm$7$\times$10$^{-3}$  &  0.2$\stackrel{+0.2}{_{-0.1}}$      &   \textbf{0.361$\pm$5$\times$10$^{-3}$}      & \\
Planet-to-star radii ratio&$R_\mathrm{p}/R_{\star}$& 0.09049$\pm$8$\times$10$^{-5}$&0.099$\pm$4$\times$10$^{-3}$&\textbf{0.09050$\pm$6$\times$10$^{-5}$} $^{\dagger}$& \\
Stellar rotation velocity & $v$sin$i_{\star}$      & 		-				 &62.9$\pm$1.4 		   &\textbf{60.0$\stackrel{\textbf{+0.9}}{_{\textbf{-0.8}}}$} 	   &	km\,s$^{-1}$ \\
Local line width 			& $s^{\dagger\dagger}$	& 		-	 &0.084$\pm$6$\times$10$^{-3}$ 	 &\textbf{0.076$\pm$5$\times$10$^{-3}$}	          &	\\
Systemic velocity 			& $\gamma_\mathrm{2012}$ & 	- 				 &-16.1$\pm$1.1					 &\textbf{-14.7$\pm$1.1} 					&	km\,s$^{-1}$ \\
Sky-projected obliquity 	& $\lambda$ 					&  - 			 &12.0$\stackrel{+13.4}{_{-6.6}}$  		 &\textbf{12.5$\stackrel{\textbf{+3.0}}{_{\textbf{-2.9}}}$} 	  &	deg\\      
\hline									
\hline
\multicolumn{6}{l}{$\dagger$: These parameters are constrained with priors from photometry. } \\
\multicolumn{6}{l}{$\dagger\dagger$: $s$=FWHM/(2\,$\sqrt{2\,ln\,2}$\,$v$sin$i_{\star}$) } \\
\end{tabular}
\label{tab_tomo}
\end{table*}

\subsection{Consistency checks}
\label{consistent}

\subsubsection{Model parameters}
\label{check_mod}
		The shape of the stellar line profile and the amplitude of the planetary signature near ingress and egress depend on the effects of limb-darkening. In the present case however, the quality of the data and the phase coverage/resolution do not allow for the determination of the limb-darkening coefficient. All model parameters values derived in Sect.~\ref{results} thus remained within their 1$\sigma$ uncertainties when $\epsilon$ was varied between 0.1 and 0.5. Because of the quality of the data, using a Voigt profile instead of a Gaussian profile to model the local line profile had no influence on the fit. \\
		We ran a tomographic fit for an eccentric orbit with $e$ set to its $3\sigma$ upper limit of 0.72 and found that all model parameters, in particular the obliquity, remained consistent within their 1$\sigma$ uncertainties with their best-fit values derived in Sect.~\ref{results}. Similar results were obtained for a circular orbit with the Keplerian semi-amplitude set to its $3\sigma$ upper limit of 0.51\,km\,s$^{-1}$ (Sect.~\ref{kep_fit}). Both eccentricity and semi-amplitude affect the shape of the Keplerian velocity curve, shifting the location of the stellar rest velocity in each of the observed CCF. This can in theory modify the best-fit values for the obliquity, systemic velocity, and stellar rotation velocity for the reasons given in Sect.~\ref{results}. Here though, RV variations at the time of the observations remained small enough to have no significant effect on the results.

\subsubsection{Run 2013}
\label{check_runB}

\begin{figure}[tbh] 
\centering
\includegraphics[trim=0cm 4cm 1.8cm 1cm,clip=true,width=\columnwidth]{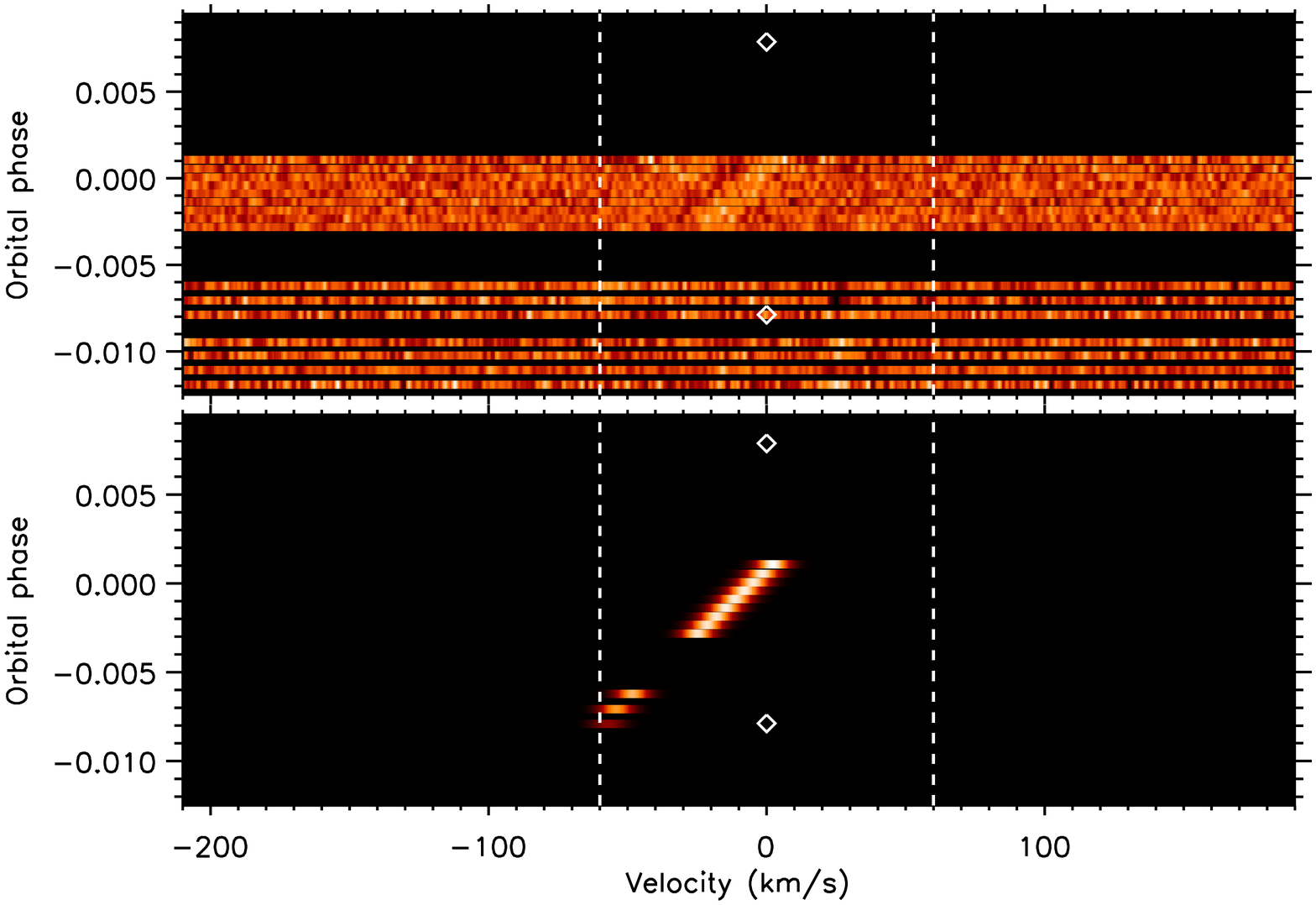}	
\includegraphics[trim=0cm 2cm 1.8cm 9.7cm,clip=true,width=\columnwidth]{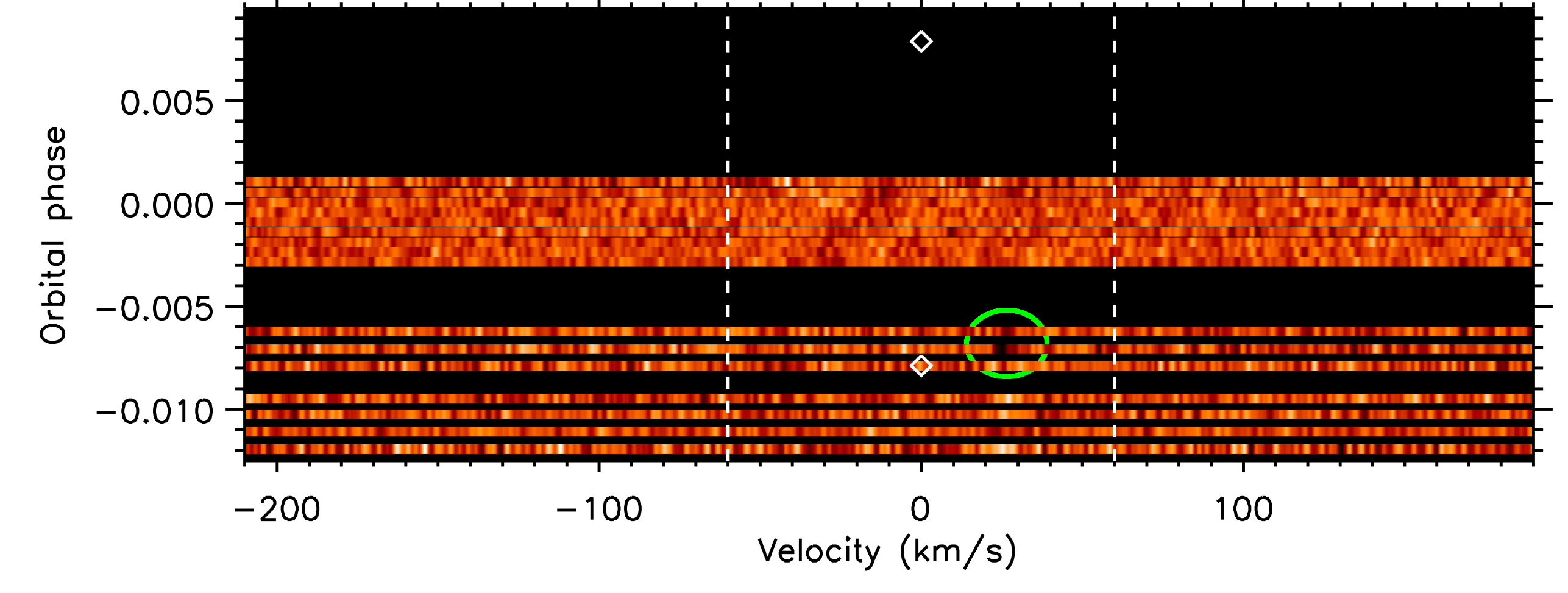}	
\caption[]{\textit{Upper panel}: Residual map of dataset 2013 after subtraction of Run 2012 best model stellar profile, plotted on the same scales as in Fig.~\ref{fig:residus_final} for a better comparison. CCFs are plotted as a function of radial velocity relative to the star, with vertical dashed white lines at $\pm$ $v$sin$i_{\star}$. Orbital phase increases vertically, with white diamonds plotted at ingress and egress. The signature of KOI-12.01 is mainly visible as a moderately bright streak near the center of the transit, and is consistent in terms of phase and velocity with the best-fit model signature from Run 2012, displayed in the middle panel at the phases of Run 2013 spectra. \textit{Lower panel}: Residual map after the further subtraction of Run 2012 model planet signature. Data is noisier at the beginning of the run, with two CCFs in particular (phases -0.0071 and -0.0062) showing sharp anomalous dips near 25\,km\,s$^{-1}$ (highlighted with a green ellipse).}
\label{fig:res_2013}
\end{figure}

We performed the tomographic analysis of dataset 2013 alone to check its consistency with dataset 2012, but found we could not obtain a good fit to the data. Instead, we calculated the residual maps for Run 2013 using the best-fit stellar line profile derived in Sect.~\ref{results} for Run 2012 (Fig.~\ref{fig:res_2013}). The transiting signature of KOI-12.01 was revealed in the residuals (upper panel), and is consistent with the best-fit planet signature from Run 2012 analysis (middle panel). \\
The fit to dataset 2013 may have failed for several reasons. Tomography is most efficient when the entire planet trajectory can be sampled in the time-series CCFs, as was the case in Run 2012. Unfortunately, Run 2013 has a poor phase coverage with most measurements taken near the center of the transit and nearly no observations during its second half (see Fig.~\ref{fig:fit_kepler}). In addition, the lack of reference spectra outside of the transit likely hindered the correction of the asymmetric continuum. The quality of the adjustment may also have been decreased by anomalies such as the low fluxes in the red wing of two CCFs near ingress (see Fig.~\ref{fig:res_2013}). These features are responsible for the spurious RV measurements mentioned in Sect.~\ref{rm_ano}, since an artificial dip in the red wing of a CCF forces its Gaussian adjustement toward a higher fitted RV. Residual maps obtained with tomography are a powerful way to identify such anomalies, especially when their influence on the RV measurements cannot be identified directly, as was the case in Fig.~\ref{fig:fit_kepler}.

\subsubsection{Tomography-only analysis}
\label{unconstr}

\begin{figure*}
\centering
\begin{minipage}[b]{0.9\textwidth}		
\includegraphics[trim=4cm 2cm 7cm 1cm, clip=true,width=\columnwidth]{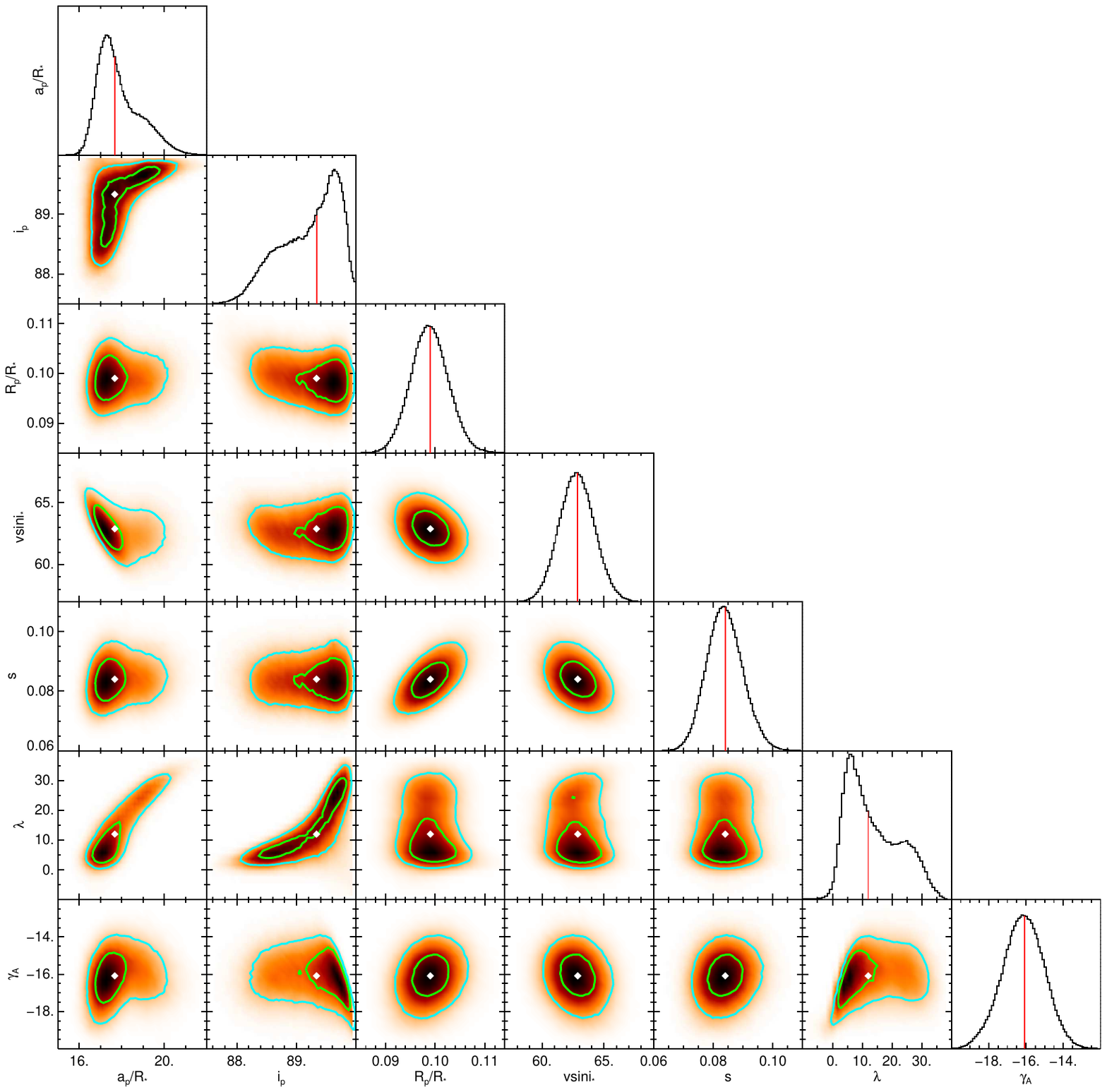}	
\end{minipage}
\caption[]{Correlation diagrams for the probability distributions of the tomographic model parameters, without priors from photometry. The plot layout is the same as in Fig.~\ref{fig:distrib_mcmc}}
\label{fig:distrib_mcmc_bis}
\end{figure*}

We investigated whether tomography alone could yield results consistent with the photometry and photometry+tomography analyses. We used the same procedure but removed the constraints from the photometric priors. The first step of the fitting procedure (Sect.~\ref{model}) showed some of the model parameters to yield a good fit to the data over a wide range of values, and we thus started the MCMC chains from diverse sets of acceptable parameter values. All chains nonetheless converged toward similar distributions, whose cumulated posterior probability distributions are displayed in Fig.~\ref{fig:distrib_mcmc_bis}. The resulting best-fit parameter values are given in Table.~\ref{tab_tomo}. \\
We sought to understand why some of the parameters present strong correlations and much larger uncertainties than in the photometry+tomography analysis. A band of stellar surface parallel to the spin axis is characterized by a unique radial velocity $v_{p}$ proportional to its distance $x_{\bot}$ to the stellar rotation axis (Eq.~\ref{eq_linvit}). Neglecting the effects of limb-darkening, the signature of KOI-12 at a given velocity $v_{p}$ in the CCFs can thus come from the occultation of any part of the stellar band at $x_{\bot}$. As many combinations of $a_\mathrm{p}/R_{\star}$, $i_\mathrm{p}$ and $\lambda$ yield the same value of $x_{\bot}$ (Eq.~\ref{eq_distaxe})\footnote{$a_\mathrm{p}/R_{\star}$ and $i_\mathrm{p}$ can be constrained independently thanks to the phase-dependent term in Eq.~\ref{eq_distaxe}}, the absence of photometry priors for these parameters result in larger uncertainties. The degeneracy is limited by the good phase coverage of the 2012 transit and corresponding good sampling of the $x_{\bot}$ values. Although the distributions for $a_\mathrm{p}/R_{\star}$, $i_\mathrm{p}$ and $\lambda$ may seem bimodal (Fig.~\ref{fig:distrib_mcmc_bis}) we found that the two apparent modes are part of a unique region of minimal $\chi^2$. There is thus only one set of acceptable values for all parameters, albeit with large uncertainties. \\
We then compared the consistency of these results with those obtained in the previous analyses. Best-fit values for $R_\mathrm{p}/R_{\star}$ and $v$sin$i_{\star}$ are higher by about 2$\sigma$ with respect to their values derived from the photometry+tomography analysis. It is unclear why we obtain this difference for the projected stellar rotation velocity, since it should be well constrained by the shape of the unocculted stellar line profile and by the spectral trajectory of KOI-12.01 in the residual map. The large value for $R_\mathrm{p}/R_{\star}$, however, may have several origins. First, macro-turbulence elongates the wings of the local stellar line profile (e.g., \citealt{hirano2011b}), affecting the width of the planet bump. Because we assumed this profile to be gaussian, and KOI-12 has a large macro-turbulence velocity ($v_{macro}\sim$18\,km\,s$^{-1}$, higher than the resolving power of the SOPHIE spectrograph at 7.5\,km\,s$^{-1}$), the fit may have been biased toward a deeper gaussian profile, ie higher values for $R_\mathrm{p}/R_{\star}$. Secondly, the width of the planet bump depends on photospheric turbulence, and can be affected by blur of the planet motion if the exposure time is too long. We found that four in-transit exposures, in the first half of Run 2012 transit, have exposure durations larger than about 20mn, which is the limit we estimated for motion blur to become significant\footnote{A rough estimate of the required exposure duration can be obtained by assuming the bump should not move by half of the local line profile width during the exposure}. It is unclear, however, if such a small number of blurred exposures would be enough to bias the estimation of $R_\mathrm{p}/R_{\star}$. A last possibility may be the merging of KOI-12.01 signature with other features near phase 0.004 (see middle panel in Fig.~\ref{fig:residus_final}). The resulting brighter region is located in the best-sampled part of the time-series spectra which constrains much of the fit, and may thus force the amplitude of the model planet signature, proportional to the planet-to-star surface ratio, toward larger value. As for the analysis of Run 2013, this illustrates how tomography depends on the phase sampling and the quality of the spectra (e.g. statistical noise, or signatures due to stellar activity), two points which can limit the ability of this technique to correctly retrieve the planet signature and the corresponding system properties. Except for $R_\mathrm{p}/R_{\star}$ and $v$sin$i_{\star}$ the tomographic analysis alone, performed on a single transit, nonetheless yields model parameter values consistent within their 1$\sigma$ uncertainties with the results from photomometry, its combination with tomography, and with the analysis of the RV anomaly for the obliquity (Table.~\ref{tab_tomo}).

\section{The planetary nature of KOI-12.01}
\label{planet}

To confirm the planetary nature of KOI-12.01, we consider here the different false positive scenarios that could have mimicked the photometric light curve observed with Kepler:
\begin{enumerate}
	\item \textit{KOI-12 and KOI-12.01 as an unblended eclipsing binary }\\
With a 3$\sigma$ upper limit on the mass of KOI-12.01 of about 10\,M$_{Jup}$, we can exclude that the photometric signature was caused by the transit of an unblended low-mass star or brown-dwarf.	\\
	
	\item\textit{KOI-12 blending the light of an eclipsing binary }\\
The transit light curve could have been produced by the occultations of an eclipsing binary in the background/foreground, diluted by the light of the fast-rotator KOI-12. This scenario can however be excluded with tomography, as the time-series spectra show that the transit signature is located within the rotationally-broadened profile of KOI-12, and correspond to the transit of KOI-12.01 in front of KOI-12. \\
	
	\item\textit{KOI-12 and KOI-12.01 as an eclipsing binary, blended with the light of a tertiary star}\\
Using PASTIS, we found that the only triple system consistent with the Kepler photometry would need KOI-12.01 to be a small star with $R_\mathrm{KOI-12.01}/R_\mathrm{KOI-12}$\,$\sim$0.1 and a mass of about 0.14$M_\mathrm{\sun}$. The high mass of KOI-12.01 would induce a large reflex motion on KOI-12, which would result in a large Doppler shift of its line profile. In this triple system, the line profile of KOI-12 is diluted by the light of the tertiary star. There are only two possibilities for the light of this star to be undetectable in the observed spectra. In the first case, the tertiary star and KOI-12 would need to have similar projected rotational velocities, radial velocities, and brightness. This case is not only highly unlikely, but the constraints on the photometry light curve from PASTIS favor the second case, with a faint tertiary star. PASTIS indicates that KOI-12 must be about five times brighter than the tertiary star, which means that the CCF would be dominated by the large rotationnally-broadened line profile of KOI-12. In this case, any faint dilution from the putative tertiary would not be able to mask large radial velocity variations of KOI-12 inferred from the CCF. Since no significant keplerian motion was detected in the data, we can exclude the triple system scenario.\\
We note that a binary system in which KOI-12.01 would still be a planet can be excluded for similar reasons: either KOI-12 and the tertiary star would have to be uncannily similar, or the tertiary star would have to be very faint and would not impact the measurements.
\end{enumerate}

We thus conclude to the planetary nature of the Kepler candidate KOI-12.01, herafter designated KOI-12b.


\section{Discussion and conclusion}
\label{conclu}

\begin{figure}
\centering
\includegraphics[trim=0.5cm 2cm 0.2cm 2cm,clip=true,width=\columnwidth]{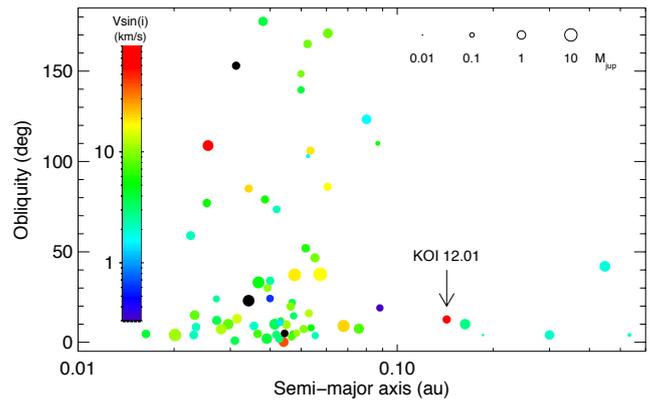}	 
\caption[]{Absolute obliquity as a function of orbital distance, for the 68 exoplanets with a measure of the spin-orbit angle. The size of each disk corresponds to the planet mass, while its color is related to $vsini_{\star}$ (color is black when the value is unknown). KOI-12b, which orbits a fast rotator at more than 0.1\,AU, is located with an arrow.} 
\label{fig:diagram}
\end{figure}

We assess the existence of a giant exoplanet around the Kepler target star KOI-12, with an upper mass limit of 10$M_\mathrm{Jup}$. The inflated radius of KOI-12b (R$_{p}$ = 1.43$\pm0.13$\,$R_\mathrm{Jup}$), unexpected for this moderately irradiated warm Jupiter (a$\sim$0.14\,au, T$_{eq}\sim$1110\,K), makes it the largest exoplanet known at orbital distance greater than 0.1\,au. This raises questions about the origin of this planetary system, which can be studied through the measurement of its obliquity. We used line-profile tomography of SOPHIE time-series spectra to identify the prograde, low-obliquity orbit of KOI-12b with $\lambda$ = 12.5$\stackrel{+3.0}{_{-2.9}}^\circ$. Doppler tomography allowed a higher precision on the obliquity to be reached by comparison with the analysis of the RM anomaly detected in the RV measurements derived from the SOPHIE spectra, although the two methods yielded consistent results. This technique can also be used to detect flux anomalies in the spectra, that can bias the RV measurements.  \\
Time-series CCFs used in tomography are generally computed using over-sampling with velocity/wavelength bins much smaller than the instrumental resolution. This results in correlated noise preventing the use of $\chi^2$ statistics to adjust the model parameters. In previous studies (e.g. \citealt{cameron2010a}; \citealt{miller2010}), uncorrelated Gaussian noise was recovered by resampling spectra at the instrumental resolution. Here we devised an empirical method, based on the analysis of the tomography residuals, to estimate statistically independent error bars on the unbinned spectra. Because of the size of KOI-12b and the very high rotational broadening of its host star, the velocity width of the missing starlight was large enough to be detected in the spectra binned at the instrumental resolution. In the future, our method should allow for smaller planets, or planets transiting stars with very low rotational broadening, to be more easily detected using tomography. \\
As can be seen in Fig.~\ref{fig:diagram}, most obliquity measurements have been made for massive exoplanets at close orbital distance and until now only five planets at more than 0.1\,au (or $P$\,$\approxsup$11\,days) had their alignment known. HD80606b and HD17156b are two giant planets on a highly eccentric orbit. The former orbits alone its host-star HD\,80606, in a binary system with HD\,80607, and shows a large misalignement (\citealt{hebrard2010}); the latter displays a low obliquity (\citealt{barbieri2009}, \citealt{narita2009}) and possibly has a second planet (\citealt{Short2008}). The last case is the Kepler-30 system, which hosts a giant planet and two smaller planets on a coplaner, well-aligned orbit (\citealt{SanchisOjeda2012}). By comparison, KOI-12b is an isolated giant planet on a slightly misaligned orbit. It is hazardous to search for trends in this limited sample, but for now obliquities of planetary systems beyond 0.1\,au seem lower and less varied than for closer-in planets (see Fig.~\ref{fig:diagram}).\\
With an effective temperature T$_{\rm eff}$=6820$\pm$120\,K and a projected rotational velocity $v$sin$i_{\star}$=60.0$\pm0.9$\,km\,s$^{-1}$, KOI-12 is one of the hottest star known to host an exoplanet and the fourth fastest rotator host after WASP-33 ($v$sin$i_{\star}$=86.1$\pm$0.4\,km\,s$^{-1}$, \citealt{cameron2010}), Kelt-7b ($v$sin$i_{\star}\sim$73$\pm0.5$\,km\,s$^{-1}$, \citealt{Bieryla2015}), and the binary system KOI-13 ($v$sin$i_{\star}\sim$70\,km\,s$^{-1}$, \citealt{Santerne2012b}). With an obliquity significantly higher than 0$^{\circ}$, KOI-12b can be considered as slightly misaligned. Yet with $\abs{\lambda}<$30$^{\circ}$ the system is not misaligned in the sense defined by \citet{Winn2010a}, and does not follow the apparent trend that misaligned planets are found around hot stars (T$_{\rm eff}>$6250\,K). \citet{Winn2010a} suggested that the thin convective envelop of a hot star limits tidal dissipation, preventing the realignement of misaligned systems. It is possible that the KOI-12 system acquired early-on a moderate obliquity, that it kept to this day because of the limited tidal damping. In addition, the large orbital distance of KOI-12b and the large mass of its host star (1.452$\pm$0.093\,M$_{sun}$) may have increased the tidal-dissipation timescale (\citealt{albrecht2012}) in this relatively young system of 1.5$\pm$0.5\,Gyr. The KOI-12 system would make an interesting target to study the link between obliquity and (elliptical) tidal instability (\citealt{cebron2011}), and with a moderately bright host-star (Kepler magnitude 11.4) KOI-12b will be a precious target for future exoplanet atmosphere investigations. \\


\begin{acknowledgements}
We would like to offer particular thanks to the referee for their in-depth reading and very constructive comments. This publication is based on observations collected with the NASA satellite Kepler and the SOPHIE spectrograph on the 1.93~m telescope at \textit{Observatoire de Haute-Provence} (CNRS), France (programs 12A.PNP.MOUT, 13A.PNP.MOUT, and 14A.PNP.HEBR). The authors acknowledge the support of the French Agence Nationale de la Recherche (ANR), under program ANR-12-BS05-0012 "Exo-Atmos". This work has also been supported by an award of the Fondation Simone et Cino Del Duca. A.S. is supported by the European Union under a Marie Curie Intra-European Fellowship for Career Development with reference FP7-PEOPLE-2013-IEF, number 627202. A.~S.~Bonomo acknowledges funding from the European Union Seventh Framework Programme (FP7/2007-2013) under Grant agreement number 313014 (ETAEARTH). This research has made use of the Extrasolar Planets Encyclopaedia at exoplanet.eu as well as the Exoplanet Orbit Database and the Exoplanet Data Explorer at exoplanets.org.
\end{acknowledgements}

\bibliographystyle{aa} 
\bibliography{biblio} 

\end{document}